%% file: libcomp.tex
\documentclass{sig-alternate} 

\usepackage{graphicx}
\usepackage{amsthm}
\usepackage{isolatin1}
\usepackage{charter}
\usepackage{ulem}
\usepackage{eprint}
\usepackage{epic}
\usepackage{eepicemu}
\usepackage{hyperref}
\usepackage{subfigure}
\input{macros}

\newcommand{\ifverbose}[1]{}
\newcommand{\ifnotverbose}[1]{#1}

\begin{document}
%

\title{Software Libraries and Their Reuse: \\ Entropy, Kolmogorov Complexity, and Zipf's Law \titlenote{Accepted for Library-Centric Software Design (LCSD'05), an OOPSLA 2005 workshop.  \href{http://arxiv.org/abs/cs.SE/0508023}{arXiv:cs.SE/0508023}}}
\subtitle{[Extended Abstract]}
%
%

\numberofauthors{1}
%

\author{
%
\alignauthor Todd L. Veldhuizen \\
       \affaddr{Open Systems Laboratory}\\
       \affaddr{Indiana University Bloomington}\\
       \affaddr{Bloomington, IN, USA}\\
       \email{tveldhui@acm.org}
}
\maketitle
\begin{abstract}
We analyze software reuse from the perspective of information theory
and Kolmogorov complexity, assessing our ability to ``compress'' programs
by expressing them in terms of software components reused from libraries.
A common theme
in the software reuse literature is that if we can only get the right
environment in place--- the right tools, the right generalizations,
economic incentives, a ``culture of reuse'' --- then reuse of software
will soar, with consequent improvements in productivity and software
quality.  The analysis developed in this paper paints a different picture:
the extent to which software reuse can occur is an intrinsic property
of a problem domain, and better tools and culture can
have only marginal impact on reuse rates if the domain is inherently
resistant to reuse.
We define an entropy parameter $H \in [0,1]$ of problem domains that measures 
program diversity, and deduce from this upper bounds on code reuse and the 
scale of components with which we may work.  For ``low entropy'' domains with
$H$ near $0$, programs are highly similar to one another and
the domain is amenable 
to the Component-Based Software Engineering (CBSE) dream of programming by 
composing large-scale components.
For problem domains with $H$ near $1$, programs require substantial
quantities of new code, with only a modest proportion of
an application comprised of reused, small-scale components.
Preliminary empirical results from Unix platforms support
some of the predictions of our model.
\ignore{***
A concrete prediction arising from our analysis is that the frequencies
with which library components are reused should follow a Zipf-style law,
confirmed by empirical measurements we make of reuse rates on several
Unix platforms.
Finally, we prove an incompleteness result for domain-specific libraries,
that there are infinitely many useful software components and no finite
library can achieve optimal reductions in program size.
***}
\end{abstract}


\terms{Software Libraries, Reuse, Entropy, Information Theory, Kolmogorov
Complexity, Zipf's Law, Component-Based Software Engineering}



\ignore{***
\section*{Todo}

Define $H$ to coincide with AEP.

Simplify the presentation of the distribution: define a notation for
all programs less than $s_0$ bits in size.

Citations for the use of a family of distributions (Levin).

What's going on with linux data?

Include tutorial on information theory

Cite better book on information theory.

Rework the ``size'' section derivation.

Get data on subroutine growth.

Prove the theorem for "Entropy maximization"

Does the existence of a mean program size imply $H=0$?

Write a better conclusion.  What does the analysis say about
the viability of strong reuse?

Heap's law: mention the role of software library evolution and
growth, ``effective $H$''.

Revised explanation for Zipf's law: (1) it maximizes entropy; (2)
if there was a significant departure from Zipf's law then the
entropy rate would be less than optimal, and therefore there would
be local exploitable patterns as per the Asymptotic Equipartition
Property.
***}

\section{Introduction and Overview}

\input{intro}

\input{summary}

\subsection{Organization}

The remainder of this paper is organized as follows.
\refsec{s:model} introduces an abstract model of software
reuse from which we derive our results.
In \refsec{s:background} we give a brief overview of 
Kolmogorov complexity.
In \refsec{s:reusebound} we derive bounds on the rates at
which software components may be reused, and give an account
for the appearance of Zipf-style empirical laws.
\refsec{s:reusepotential} examines the potential for software
reuse as a function of the parameter $\Hdomain$.
In \refsec{s:data} we present some preliminary experimental
results, and \refsec{s:conclusion} concludes.

\section{Modelling library reuse}
\input{model}

\section{Kolmogorov Complexity}

\label{s:background}

\input{kolmogorov}

\section{A bound on reuse rates}

\label{s:reusebound}
In this section we derive a bound
on the reuse rate $\lambda(n)$ at which the $n^\mathit{th}$ library component
is reused in `compressed' programs written with use of a library.

\input{coding}

\input{newderiv}

\section{Reuse potential}

\label{s:reusepotential}
In the following sections we consider the possibilities of
code reuse in two cases: (1) when $\Hdomain = 1$ and we
have a uniform distribution on programs; (2)
when $0 < \Hdomain < 1$ and we have some degree of
compressibility in the problem domain.
The case $\Hdomain=0$ is left for future work.

\input{uniform}
\input{nonuniform}
\input{size}

\input{data}
\input{conclusion}

\section{Acknowledgments}

This paper benefited immeasurably from discussions with my colleagues
at Indiana University Bloomington.  In particular I thank
Andrew Lumsdaine, Chris Mueller, Jeremy Siek, Jeremiah Willcock, 
Douglas Gregor, Matthew Liggett, Kyle Ross, and Brian Barrett for their 
valuable suggestions.  
I thank Harald Hammerstr{\"o}m for letting me disappear with
his copy of Li and Vitányi \cite{Li:1997} for most of a year.

\bibliographystyle{plain}
\input{biblio}

\appendix
\section{Background}
\input{asymptotics}


\end{document}

%% file: macros.tex
\newcommand{\editor}[1]{\begin{sc}#1\end{sc}}

\newtheorem{thm}{Theorem}[section]
\newtheorem{lem}{Lemma}[section]
\newtheorem{prop}{Proposition}[section]
\newtheorem{claim}{Claim}[section]
\newtheorem{cor}{Corollary}[section]
\newtheorem{fact}{Fact}[section]
\newtheorem{assume}{Assumption}
\newtheorem{hypothesis}{Hypothesis}
\theoremstyle{definition}
\newtheorem{defn}{Definition}
\newtheorem{principle}{Principle}

\newcommand{\reffig}[1]{Figure~\ref{#1}}
\newcommand{\refsec}[1]{Section~\ref{#1}}
\newcommand{\reflem}[1]{Lemma~\ref{#1}}
\newcommand{\refthm}[1]{Theorem~\ref{#1}}
\newcommand{\refcor}[1]{Corollary~\ref{#1}}
\newcommand{\reffact}[1]{Fact~\ref{#1}}
\newcommand{\refclaim}[1]{Claim~\ref{#1}}
\newcommand{\refprin}[1]{Principle~\ref{#1}}
\newcommand{\refprop}[1]{Proposition~\ref{#1}}
\newcommand{\refeqn}[1]{Eqn.\ (\ref{#1})}
\newcommand{\refineqn}[1]{Inequality (\ref{#1})}
\newcommand{\refdefn}[1]{Defn.\ \ref{#1}}

\newcommand{\refassume}[1]{Assumption \ref{#1}}

\newcommand{\ignore}[1]{}
\DeclareSymbolFont{AMSb}{U}{msb}{m}{n}
\DeclareMathSymbol{\N}{\mathbin}{AMSb}{"4E}
\DeclareMathSymbol{\Z}{\mathbin}{AMSb}{"5A}
\DeclareMathSymbol{\R}{\mathbin}{AMSb}{"52}
\DeclareMathSymbol{\Q}{\mathbin}{AMSb}{"51}
\DeclareMathSymbol{\I}{\mathbin}{AMSb}{"49}
\DeclareMathSymbol{\C}{\mathbin}{AMSb}{"43}

\newcommand{\paragraphlabel}[1]{}

\newcommand{\tinysection}[1]{\vspace{0.1in}\noindent{\bf #1}}

\newcommand{\strlen}[1]{\| #1 \|}  
\newcommand{\Hdomain}{H}

%% file: intro.tex

\label{s:intro}
Software reuse offers the hope that software construction can be made
easier by systematic reuse of well-engineered components.  
\ifverbose{The
importance of developing libraries of reusable components has
been recognized since the very earliest days of computing (e.g.,
\cite{Wheeler:ACM:1952}), but received  a substantial push with 
M. Douglas McIlroy's famous paper \emph{Mass Produced Software Components}
\cite{McIlroy:NATO:1969}.  }
In practice reuse has been found to
improve productivity and reduce defects
\cite{Basili:CACM:1996,Griss:IBMSJ:1993,Krueger:CSUR:1992,Mohagheghi:ICSE:2004,Poulin:1997}.
But what of the limits of reuse --- will large-scale reuse make software 
construction easier?  Thinking here is varied, but for the sake of
argument let me artificially divide the opinions into two competing 
hypotheses.  First the more enthusiastic end of the spectrum, which I associate
with the Component-Based Software Engineering (CBSE) movement.

\begin{hypothesis}[Strong reuse]
Large-scale reuse will allow mass-production
of software, with applications being assembled by composing
large, pre-existing components.  The activity of programming will consist
primarily of choosing appropriate components from libraries, adapting and
connecting them.
\end{hypothesis}

Strong reuse is thought to thrive in problem domains with
great concentration of effort and similarity of purpose, i.e.,
many people writing similar software whose requirements show
only minor variation.
\ifverbose{
One offshoot
of ``strong reuse'' thinking has been the development of special-purpose
languages for describing the composition of components, variously
called composition languages \cite{Achermann:FMDP:2001}, module
interconnection languages \cite{Prieto-Diaz:JSS:1986}, coordination
languages \cite{Gelernter:CACM:1992}, configuration languages
\cite{Bishop:ICCDS:1996}, or simply scripts \cite{Schneider:1999}.
A hallmark of such languages is that they are not intended for
general-purpose programming, but specialized for describing
how pre-existing components are wired together.
}
However, the question of
whether strong reuse can succeed for software construction considered
globally, across disciplines and organizations, remains uncertain.
A more cautious view of reuse is the following.

\begin{hypothesis}[Weak reuse]
Large-scale reuse will offer useful
reductions in the effort of implementing software, but these savings
will be a fraction of the code required for large projects.  Nontrivial
projects will always require the creation of substantial quantities of new
code that cannot be found in existing component libraries.  
\end{hypothesis}

Representative of weak reuse thinking is the following prescription
for code reuse in well-engineered software from Jeffrey Poulin 
\cite{Poulin:1997}:
up to 85\% of code ought be reused from libraries,
with a remaining 15\% custom code, written specifically for
the application and having little reuse potential.
\ifverbose{
Weak reuse is most likely to succeed in problem domains where there
are substantial numbers of people writing code,
and there is enough commonality between the software projects undertaken
to support the development of libraries for the problem domain.
Weak reuse is also thought to
work in organizations that produce many software products with greatly
overlapping functionality --- the impetus behind the ``Software Product
Lines'' movement, e.g., \cite{Clements:2001}. 
}
The percentage of code that may be reused from libraries
varies greatly across problem domains, but
weak reuse paints a fairly accurate
picture of the software landscape of today.
Many explanations
are proposed for why strong reuse is not happening on a global
scale (cf. \cite{Frakes:SE:1996}).  A common position in the reuse
literature is that if we can only get the right environment in place --- the
right tools, generalizations, economics, a ``culture of reuse'' --- 
then reuse of software will soar, with consequent improvements in productivity
and software quality.

\tinysection{A contrary view.}
The perspective developed in this paper
suggests that the extent to which reuse can happen is an intrinsic property
of a problem domain, and that improving the
ability of programmers to find, adapt, deploy, generalize and market components
will have only marginal impact on reuse rates if the domain is 
resistant to reuse.
We propose to associate with problem domains an entropy parameter 
$0 \leq \Hdomain \leq 1$
measuring the diversity of a problem domain.
When $\Hdomain=1$, software is
extremely diverse and we should expect very little potential
for reuse; in fact, we show that the proportion of an application
we can draw from libraries approaches zero for large projects.  For
problem domains with $\Hdomain < 1$, software is somewhat homogeneous,
and with decreasing $\Hdomain$ comes increasing potential for reuse.
The theory we develop suggests that an expected proportion of \emph{at most}
$(1-\Hdomain)$ of an application's code may be reused from libraries,
with a remaining proportion $\Hdomain$ being custom code written
specifically for the application.  As $\Hdomain$ nears $0$ we enter
the strong reuse utopia of ``programming by composing large
components.''  The possibilities of reuse are 
\emph{strictly limited} by the parameter $\Hdomain$, which is an intrinsic 
property of the problem domain.

\ignore{***
A cynical view of reuse is that the availability of vast libraries
of reliable components will succeed in making us more productive, but
at the same time make the activity of programming increasingly difficult.
Lisanne Bainbridge, a psychologist studying industrial automation,
wrote a wonderful paper \emph{Ironies of Automation} \cite{Bainbridge:NTHE:1987}.  She observed that automation often makes an operator's job harder, 
in part because the tasks easy to automate are largely those easy for the
operator to perform, and their
removal may leave the operator with an arbitrary collection of
difficult tasks.  Thomas Green suggested this style of cautionary
thinking might apply to software also, and we should beware ``Ironies
of Abstraction.'' \cite{Green:unpub:...}
Perhaps large-scale reuse will take away the simple parts of programming,
and we shall have to think very hard about everything we do.

But cautionary tales aside, let us return to this question of
strong reuse.  
***}

We develop this theory by examining
our ability to compress or compactify software by the use
of libraries.
We shall speak throughout this paper of \emph{compressed programs}, by
which we mean programs written using libraries, and \emph{uncompressed
programs} that are stand-alone and do not refer to library components.
The principle tools we employ 
are information theory and Kolmogorov complexity.
Both of these carry subtly different notions of compressibility that
we shall have to juggle.
The information theory notion deals with
compressing objects by identifying 
patterns that appear frequently and giving them short descriptions
--- as in English we have taken
to saying ``car'' for ``automobile carriage.''  The Kolmogorov 
version of compressibility describes our ability to find for a
given program a shorter program with the same behaviour, without
appealing to how typical that program might be
for the problem domain within which we are working.
We assume some basic familiarity with information theory as 
might be found in e.g. \cite[Ch. 2]{Cover:1991} or \cite{Li:1997}.
The essentials of Kolmogorov complexity are reviewed in
\refsec{s:kolmogorov}.

\tinysection{Library components and prime numbers.}
Integers factor into a product of primes;
software can be factored into an assembly of components.
Library components are the prime numbers of software.
This would be a terribly naive thing to say were it not for the many 
wonderful parallels that turn up:
\begin{itemize}
\item There are infinitely many primes; in \refsec{s:incompleteness}
we prove there are infinitely many components for a problem domain that
reduce expected program size (thus guaranteeing employment for
library writers.)
\item The $n^\mathit{th}$ prime is a factor of $\sim \frac{1}{n \ln n}$
of the integers.  Theory predicts the $n^\mathit{th}$ most
frequently used library component has an ideal reuse rate of about
$\frac{1}{n \log n \log^+ n}$ (\refsec{s:newderiv}).
\item The Erd{\"o}s-Kac theorem states that the number of factors of
an integer tends to a normal distribution; we measure experimental
data that suggests a similar theorem might be provable for software
components (\reffig{f:erdos}).
\item The Prime Number Theorem states that the $n^\mathit{th}$
prime is $\sim \log (n \ln n)$ bits long.  We show that the ideal
configuration for libraries is that the $n^\mathit{th}$
most frequently used component is of size $\geq \log n$ and
$\leq \frac{1-\Hdomain}{\Hdomain} o(n^\epsilon)$ for $\epsilon \geq 0$
(\refsec{s:size}).
\end{itemize}
\ignore{***
There are limits to the analogy ---
unique factorization
of software into components would contradict undecidability of program
equivalence, for example \editor{Only if you assumed an effective procedure
for such a factorization?} --- but the similarities are
intriguing.
\editor{Would it contradict index theorems in computability?  with
an infinite undecidable library?}
***}

\input{pyramid}

%% file: pyramid.tex
\tinysection{Reuse and Zipf's Law.}
\ifverbose{
\begin{figure}
\begin{centering}
\includegraphics[scale=0.5]{pyramid}

\caption{\label{f:pyramid}An illustration of where software
abstractions are provided.  The few abstractions needed very
frequently are provided in hardware or by programming languages; the vast 
majority of abstractions are provided by software libraries.}
\end{centering}
\end{figure}
}\ifverbose{
Useful software abstractions are provided in three basic ways:
in hardware, in programming languages, or in libraries.
The level at which abstractions are realized is closely related
to how frequently they are needed.
Integer and floating-point arithmetic are provided in hardware, since 
almost every computer user runs software requiring these.
\ifverbose{  Sine and cosine 
functions are needed in a variety of applications from scientific
computing to gaming, so these are sometimes realized in hardware also.
}
Some common abstractions that don't merit machine instructions are provided
by programming languages: for example, complex numbers and string
manipulation are often provided by languages.
Less frequently used abstractions are relegated to software libraries,
and these constitute the bulk of reusable abstractions.  We can think of
abstractions as forming an iceberg, with the tip 
being those abstractions provided in hardware and languages, and the vast
majority lurking in libraries\ifverbose{ (\reffig{f:pyramid})}.
In our view, machine instructions are a special case of ``software components.''
}
\begin{figure}
\begin{centering}
\includegraphics[scale=0.5]{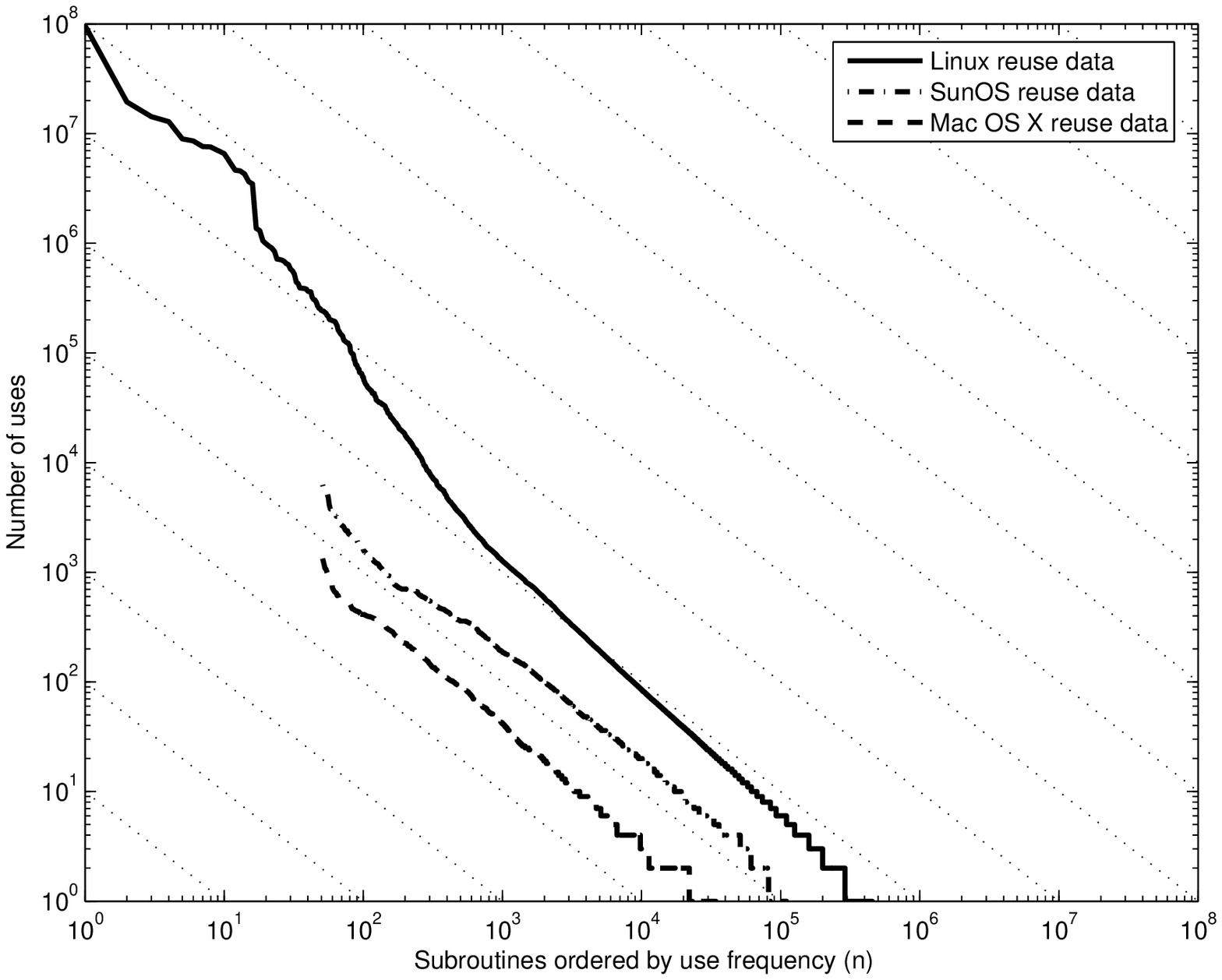}

\end{centering}
\caption{\label{f:rates}Data collected from shared objects on
several unix platforms, showing the number of references to library
subroutines.  The observed number of references shows
good agreement with Zipf-style frequency laws of the form $c \cdot n^{-1}$
(dotted diagonal lines).
A detailed explanation of this data is given in \refsec{s:data}.}
\end{figure}
It is known that hardware instruction frequencies follow an
iconic curve described by George K. Zipf for
word use in natural languages \cite{Kuck:1978,Latendresse:IVME:2003,Wortman:1973}.
Zipf noted that if words in a natural language are ranked
according to use frequency, the frequency of the $n^{\mathit{th}}$
word is about $n^{-1}$.  Zipf-style empirical laws crop up in
many fields \cite{Powers:ACL:1998,Li:WWW:2005}.
Evidence suggests programming language constructs also
follow a Zipf-like law \cite{Berry:SIGPLAN:1983,Laemmel:TR:1977}.  
It is natural then to wonder if
this result might extend to library
components.  Our results support this conclusion.
\reffig{f:rates} shows the reuse counts of
subroutines in shared objects on three Unix platforms,
clearly showing Zipf-like $n^{-1}$ curves.
These results are described in detail in \refsec{s:data}.
The appearance of such curves is not happenstance.
In \refsec{s:newderiv} we argue they
are a direct result of programmers trying to write as little
code as possible by reusing library subroutines; this drives
reuse rates toward a ``maximum entropy'' configuration, namely
a Zipf's law curve.

%% file: summary.tex
\ifverbose{
\subsection{Contributions}

The major results of this paper are as follows.
\begin{enumerate}
\item We give empirical measurements showing strong evidence that
reuse of subroutines from software libraries follows a Zipf-like
distribution.
\item We prove that under reasonable assumptions the reuse
rates $\lambda(n)$ of software components (of any sort, not just
subroutines) are asymptotically bounded by $(n \log_2 n)^{-1}$.
\item We investigate the proportion of code that one might 
expect to reuse from libraries.  Adopting ideas from Kolmogorov
complexity, we associate with a given problem domain a
\emph{randomness deficit rate} that measures how diverse
software is within the problem domain.  This randomness
deficit governs how much library code we can expect to
reuse in implementing programs.  If software within a
domain is highly diverse, libraries are of little use:
asymptotically the number of components we may usefully
reuse is bounded by a constant, and the fraction of code
we can reuse from libraries tends to zero as program
size increases.  On the other hand if software is not
very diverse,  \editor{...}
\end{enumerate}
}

%% file: model.tex
\label{s:model}
In this section we propose an abstract model capturing some
essential aspects of software reuse within a problem domain.
The basic scenario is this:
we have a library, possibly many libraries that we collectively
consider as one, that contains a great number of software components.
These components may be subroutines, architectural skeletons, design
patterns, generics, component generators, or whatever form of abstraction we
may yet invent; their precise nature is unimportant for the argument.
In using a component from the library we achieve some reduction in
the size of the program, and perhaps consequently, in the effort
required to implement it.  
Program size serves as a rough lower
bound to effort, but it would be a grave error to confuse the two.

\subsection{Distribution of programs in a domain}

\label{s:distribution}
We presume that the projects undertaken by programmers working in a
problem domain can be modelled by
a probability distribution on programs.
The probability distribution is defined on ``uncompressed'' programs
that do not use any library components.
These uncompressed programs can be viewed as \emph{specifications}
that programmers set out to realize.

We consider compiled programs modelled by binary strings 
on the alphabet $\{0,1\}$.
We write $\strlen{w}$ for the length of a string $w$.\ignore{\footnote{
The use of $|w|$ for string length, while traditional in some quarters,
leads to confusing notations such as $P(w ~|~ |w| < m)$ for 
probability conditioned on string length.  We use
$\strlen{\cdot}$ to avoid this.
}}
Finite programs are countably infinite in number, so we immediately encounter
the problem of defining a probability distribution in which the
probability of encountering individual programs may be infinitesimal.
A rigorous approach would be to employ measure theory,
for example Loeb measure, which would allow us to speak of the probability
of individual programs.
This would require some rather daunting
machinery and we instead settle for a more accessible
approach similar to that used by 
\cite{Ben-David:JCSS:1992,Compton:NATO:1988,Rissanen:1989}.

Let $A^{\leq n} = \{ w \in \{0,1\}^\ast ~:~ \strlen{w} \leq n \}$ denote
compiled programs of length at most $n$ bits.
We introduce a family of conditional distributions
$\{p_{s_0}\}_{s_0 \in \N}$ whose domains consist of programs
$\leq s_0$ bits in size, that is,
\begin{align*}
p_{s_0} : A^{\leq s_0} \rightarrow \R
\end{align*}
and satisfying $\sum p_{s_0} = 1$ and $p_{s_0}(w) \geq 0$.
The intent is that $p_{s_0}(w)$ gives the
probability that someone working in the problem
domain will set out to realize the particular (uncompressed) program $w$, 
given that $w$ is at most $s_0$ bits long.
For this family of distributions to be compatible with one another we 
require that 
$p_{s_0}(w) = p_{s_0+1}(w ~|~ \strlen{w} \leq s_0)$,
i.e., we can get the distribution on length $\leq s_0$ programs by
taking a conditional probability on the distribution for length
$s_0+1$ programs.
We do not presume that such distributions can be effectively
described.

\ifverbose{
One misgiving we might have is that programmers do not
set out to duplicate a given program, but rather to realize
some specification or behaviour.  However, it turns out that
almost every compiled program is the smallest program with
its behaviour (\refsec{s:kolmogorov}),
so programs are a reasonable stand-in for specifications.
}

In what follows we use the usual notation for expectation
with the implied assumption of $s_0 \rightarrow \infty$; for
example, if $f : \{0,1\}^\ast \rightarrow \R$ maps programs to
real numbers, then by $E[f(w)]$ we mean:
\begin{align*}
E\bigl[ f(w) \bigr] &\equiv \lim_{s_0\rightarrow \infty} \sum_{w : \strlen{w} \leq s_0} f(w) p_{s_0}(w)
\end{align*}
if such a limit should exist.
For example, a mean program size $E\bigl[\strlen{w}\bigr]$ may exist for a
problem domain, but we do not require nor expect this.

\subsection{The entropy parameter $\Hdomain$}

\label{s:ergodic}
\label{s:diversity}
A key, perhaps defining, feature of a problem domain is
that there is similarity of purpose in the programs people write.
We do not expect the distribution of programs written in a
problem domain to be uniform over all possible programs, 
but rather concentrated on programs that solve certain classes
of problems typical for the domain.
We formalize this intuition by introducing a parameter
$\Hdomain$ for problem domains
measuring how far their probability distribution departs from 
uniform.  This $H$ is very similar to
entropy rate from information theory, and coincides if we are willing to
assume programs are drawn from a stationary stochastic process.
When $\Hdomain=1$
the distribution over programs is uniform, modelling extreme diversity
of software, with little opportunity for reuse.
For $\Hdomain < 1$ there is some potential for reuse.
In fact as we shall see shortly, we may expect that
up to a proportion $1-\Hdomain$ of programs may be reused
from libraries.

Define the entropy of
each distribution $p_{s_0}$ in the standard way
(see, e.g., \cite{Cover:1991,Li:1997}):
\begin{align*}
H(p_{s_0}) &= \sum_{w : \strlen{w} \leq s_0} - p_{s_0}(w) \log_2 p_{s_0}(w)
\end{align*}
This is the expected number of bits required
to represent a program of size $\leq s_0$ in this domain.
We are interested in the limit behaviour of
$\frac{1}{|A^{\leq s_0}|} H(p_{s_0})$, 
akin to the entropy rate of a random process. 
In general this limit may not exist --- there might be 
oscillations --- so we need some weaker notion of
limit.  We settle for a limsup, which gives an almost sure
upper bound on the limit behaviour.

\begin{defn}[Entropy parameter]
Define the \emph{entropy parameter} $\Hdomain$ of a problem domain
to be the greatest value that
$\frac{1}{|A^{\leq s_0}|} H(p_{s_0})$ attains infinitely
often as $s_0 \rightarrow \infty$:
\begin{align*}
\Hdomain &= \limsup_{s_0 \rightarrow \infty} \left( \frac{1}{|A^{\leq s_0}|} H(p_{s_0}) \right)
\end{align*}
\ignore{\footnote{We do not assume $\lim_{s_0 \rightarrow \infty} \frac{1}{s_0} H(p_{s_0})$
exists; there might be oscillations in
$H(p_{s_0})$.
This definition of $\Hdomain$ differs slightly from the usual definition
of entropy rate from information theory, in which the entropy rate of
infinite streams or random processes are considered.  But the two
can be considered equivalent for most purposes as $s_0 \rightarrow \infty$.
}}
As a consequence of this definition we are guaranteed
that $H(p_{s_0}) \leq s_0 H$ almost surely as $s_0 \rightarrow \infty$.
\end{defn}

We cannot hope to calculate $\Hdomain$ from first principles
except for toy scenarios, but there is hope we might estimate it
empirically.
We introduce $\Hdomain$ primarily as a theoretical tool
to model problem domains in which
people have great similarity of purpose
($\Hdomain \rightarrow 0$) or diffuse interests 
($\Hdomain \rightarrow 1$).  The main impact of $\Hdomain$
is the following.

\begin{figure}
\begin{centering}
\input{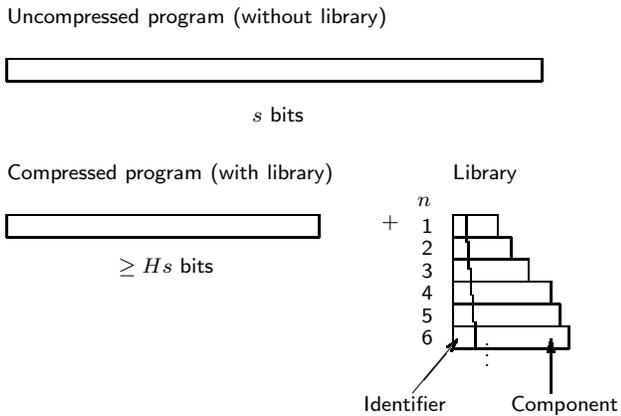}

\caption{
\label{f:Hdomain}The basic scenario: programmers in a
problem domain set out to realize a program that can
be represented in $s$ bits when compiled without the use of a library.
By using library components, they are able to reduce the
size of the compiled program, down to an expected size of
$\geq \Hdomain s$ bits.
}
\end{centering}
\end{figure}

\begin{claim}
In a problem domain with entropy parameter $\Hdomain$,
the expected proportion of code that may be reused from a library is at
most $1-\Hdomain$.
\end{claim}

This is a consequence of the Noiseless Coding Theorem of information
theory (e.g., \cite[\S 2.5]{Ash:1967}),
which states that coding random data with entropy $H$ requires
(on average) at least $H$ bits.  
Suppose an uncompressed program has size $s \leq s_0$.  
We defined $\Hdomain$ so that
$H(p_{s_0}) \leq s \Hdomain$ 
almost surely, so we can compress programs to an expected size of at best
$s \Hdomain$ by the Noiseless Coding Theorem.  
Therefore the expected amount of code saved by use of the library is at most
$(1 - \Hdomain) s$, and it is reasonable to equate this with
the amount of code reused from the library.
An immediate implication is that blanket reuse prescriptions
such as ``effective organizations reuse 70\% of
their code from libraries'' are unrealistic; reuse goals need to be pegged
to the problem domain's value of $\Hdomain$.

\reffig{f:Hdomain} illustrates the scenario we consider in
this paper: programmers set out to implement the capabilities of
some uncompressed program of length $s$
written without use of a library, drawn from the distribution for the 
problem domain.  A programmer implements the program making use of
the library, effectively ``compressing'' it.  The expected size
of the compressed program is at least $\Hdomain s$ bits, by the
previous arguments.  The library consists of a set of components,
each with an identifier or \emph{codeword} by which they are referred to.
We always take programs to be \emph{compiled}, so as not to care
about the high compressibility of source representations.


\subsubsection{Motifs and the AEP}
\label{s:AEP}
One question we should like to answer is whether when $H < 1$
there are commonly occurring patterns or ``motifs'' in programs
that we can put in libraries and reuse to compress programs.
If we are willing to assume that programs in
a problem domain behave as if excerpted from a stationary ergodic source, then
the Shannon-McMillan-Breiman theorem (asymptotic equipartition property or
AEP) \cite[\S 15.7]{Cover:1991}
ensures that when $H < 1$ there are commonly occurring
finite subsequences in programs that can be exploited, and indeed
that we can achieve optimal compression of programs merely by having
libraries of common instruction sequences.
That more complex software components
prove necessary in practice suggests the stationary ergodic 
assumption is too strong,
and a weaker ergodic property 
is needed to account for the emergence 
of motifs in software when $H < 1$.
It is unclear yet exactly what this property might be; in
the remainder of this paper we do not assume AEP.

\subsection{Libraries maximize entropy}

\label{s:entropy}
\begin{quote}
A truly great computer programmer is lazy, impatient and full of hubris. Laziness drives one to work very hard to avoid future work for a future self. --- Larry Wall
\end{quote}

Programmers, so we read, are lazy--- they write libraries to capture
commonly occurring abstractions so they do not have to write them over and
over again. 
The social processes that drive programmers to develop
libraries have an interesting theoretical effect.  We can view
programmers contributing to domain-specific libraries as collectively 
defining a system 
for \emph{compressing programs} in that domain.  If there is a common pattern,
eventually someone will identify it and put it in a library.
Since the absence of common patterns in code is implied by
high entropy, we propose the following principle.

\begin{principle}[Entropy maximization]
\label{prin:entropy}
Programmers develop domain-specific libraries that minimize the
amount of frequently rewritten code for 
the problem domain.   This tends to maximize the entropy of 
compiled programs that use libraries.
\end{principle}

As evidence for this principle, we show in \refsec{s:data}
that the rate at which library components are reused is 
empirically observed to approach a maximum entropy configuration.\footnote{
Note that \refprin{prin:entropy} is \emph{not} intended to appeal to the
maximum entropy principle as advocated by Jaynes, which deals with
maintaining uncertainty in inference.
}

In practice programmers have to strike a balance between
the succinctness of their programs and their readability; see, e.g.,
\cite{Green:HCI:1989} for an elegant discussion of such tradeoffs.
However, we maintain that the drive toward terseness and factoring
common patterns
is a defining pressure on library development: entropy is essentially
a measure of \emph{communication efficiency}, and programmers edge
as close to maximum entropy as they can while maintaining 
source-code understandability.\footnote{We re-emphasize that we are speaking of the
entropy rate of \emph{compiled} programs; source representations
are highly compressible to support readability.}

\subsection{The Platonic library}

\label{s:platonic}
In the early days of computing
libraries held a hundred subroutines at most; these days it is
common for computers to have a hundred thousand subroutines available
for reuse (cf. \refsec{s:data}).  Let us suppose that as time goes on we shall
continue to add components to our libraries as we discover
useful abstractions and algorithms.
Our current libraries might be viewed as a truncated version of some
infinite (but countable) 
library toward which we are slowly converging.  It is convenient
to pretend
that this limit already exists as some infinite ``Platonic library'' 
for the problem domain,
and that we are merely discovering ever-larger fragments of it, recalling
Erd{\"o}s' book of divine mathematical proofs.\footnote{A Platonic object
is an abstract entity thought to dwell in some realm outside 
spacetime.
Our stance with respect to software libraries echoes
mathematical Platonism, that mathematical objects about which we reason
exist in some idealized form outside the physical universe (see, 
e.g., \cite{Balaguer:SEP:2004}).
\ifverbose{Regarding Erd{\"o}s' idea: ``God has a transfinite Book, which
contains all the theorems and their best proofs, and if He is well
intentioned toward those [mathematicians], He shows them the Book
for a moment.  And you wouldn't even have to believe in God, but
you must believe that the Book exists.'' } }
Were we granted access
to the entire library, we might write software in a very efficient
way.  We use the Platonic library as a device --- a convenient
fiction --- to reason about how useful finite libraries might be.

Infinite objects need to be treated with care.
We shall not assume that some ``optimal
infinite library'' exists that is the best possible such
library.  Nor shall we assume there is some finite description or
computable enumeration of its contents.
We merely assume that fragments of the Platonic library
give us snapshots of what shall be in our software libraries over time.

\subsection{Existence of reuse rates}

\label{s:rates}
Numerous metrics have been proposed for measuring reuse.
We focus on the \emph{reuse rate} of a component, which we write
$\lambda(n)$ and define as the expected rate at which
references are made to the $n^\mathit{th}$ library component in
a compressed program.
The units of $\lambda(n)$ are expected references per bit of compiled code.
We assume mean reuse rates exist in a problem domain,
in the following sense.


\begin{assume}
\label{ass:meanlambda}
Let $\mathrm{Refs}_n(w)$ count the number of references
to the $n^\mathit{th}$ component in a compressed program
$w$ of size $\leq s_0$.  We assume that
\begin{align}
E\Bigl[\mathrm{Refs}_n(w) ~\Bigm|~ \strlen{w} = s \Bigr] ~~\sim~~ \lambda(n) s + o(s)
 ~~~~\mathrm{as}~s_0 \rightarrow \infty
\end{align}
where $o(s)$ denotes some error term growing asymptotically
slower than $s$.
\end{assume}

We unfortunately do not have a good sense of how to go from
the problem domain's distributions $p_{s_0}$ on \emph{uncompressed}
programs to rates of components in \emph{compressed} programs;
this is tied up with the ergodic process issues mentioned in
\refsec{s:ergodic}.
We dodge the issue by simply \emph{assuming} that the mean 
rates $\lambda(n)$ exist.
This is not a demanding assumption; many sensible
random process models would imply \refassume{ass:meanlambda},
for example modelling component uses as a renewal process
(see, e.g., \cite[\S 3]{Ross:1996}).\footnote{
For readers familiar with coding theory we
forestall confusion by mentioning that the rates $\lambda(n)$ are
not the same as the usual notion of probabilities over countable
alphabets.  The rates $\lambda(n)$ are 
drawn from compressed programs and so already incorporate code lengths.
}

\subsection{Ordering of library components}

\label{s:ordering}
For convenience we shall suppose the library components are arranged
in decreasing order of expected reuse rate in the problem domain: that is,
\begin{align*}
\lambda(n) \geq \lambda(n+1)
\end{align*}

There are two reasons for this.  The first is tidiness,
so that when we plot
$\lambda(n)$ vs $n$ we see a monotone function and not noise.  
The asymptotic bounds we derive on $\lambda(n)$ do not rely on this 
ordering.  The second reason is that to derive bounds on how well we
might compress programs
we need to assign shorter identifiers to more frequently used components.
This is easiest to reason about if the Platonic library is
sorted by use frequency.\footnote{Jeremiah Willcock made the useful suggestion 
that we may regard the Platonic library as containing already every possible 
component, and the only question is the order in which they
are placed.}

%% file: kolmogorov.tex

\label{s:kolmogorov}
Kolmogorov complexity, also known as Algorithmic Information Theory, was
founded in the 1960s by R. Solomonoff, G. Chaitin, and A.N. Kolmogorov.
We shall only make use of some basic facts; for a more thorough
introduction the survey article \cite{Li:HTCS:1990} or the book
\cite{Li:1997} are recommended.  The central idea is simple: measure
the `complexity' of an object by the length of the smallest program that
generates it.  This generalizes to the study of \emph{description
systems}, that is, systems by which we define or describe objects,
of which programming languages, logics, and descriptive set theory
are prominent examples.
The source code of a program, for example, describes a program behaviour;
a set of axioms describes a class of mathematical structures.  In the
general case we have some objects we wish to describe, and a
description system $\phi$ that maps from a description $w$ (for us,
a program) to objects.
The usual situation is to describe an object by exhibiting a
program that generates it; in this case we may also provide some inputs
to the program, which we shall call \emph{parameters}.  The Kolmogorov
complexity of an object $x$ in the description system $\phi$, relative
to a parameter $y$ is defined by:
\begin{align}
C_\phi(x ~|~ y) = \min_w \{ \strlen{w} : \phi_w(y) = x \}  \label{eqn:relativek}
\end{align}
In the case where the description system $\phi$ is a programming language,
we may read \refeqn{eqn:relativek} as finding the shortest program that,
given input parameter $y$, outputs $x$.  The parameter $y$ does not contribute
to the measured description length $C_\phi(x ~|~ y)$.
Without a parameter we have
the simpler case $C_\phi(x) = C_\phi(x ~|~ \epsilon)$ where $\epsilon$
is the empty string.

For example, we might choose the programming language Java as our
description system; then for some string $x$, its Kolmogorov complexity
$C_\mathsf{Java}(x)$ is the length of the shortest program that outputs 
$x$.
To determine whether use of a library $\mathsf{L}$ offers a reduction in
program size, we can consider the combination of Java and the library
$\mathsf{L}$ as a description system itself which we might call
$\mathsf{Java}+\mathsf{L}$, and
compare $C_\mathsf{Java + L}(x)$ to $C_\mathsf{Java}(x)$.

A very useful insight is that the choice of language doesn't
much matter.

\begin{fact}[Invariance {\cite[\S 2.1]{Li:1997}}]
\label{fact:invariance}
There exists a universal machine $U$ such that
if $\phi$ is some
effective description system (e.g., a programming language)
then there is a constant $c$ such that $C_U(x) \leq C_\phi(x) + c$ for any $x$.
\end{fact}
That is, the universal machine $U$ is optimal up to a constant factor.
For this reason the subscript $U$
can be dropped and one can write $C(x)$ for the Kolmogorov
complexity of $x$, knowing it is only defined up to some constant
factor.\footnote{There is an easy way to see why this is true:
if $\phi$ is a programming language, then we can write a
$\phi$-interpreter for the universal machine $U$.  We can
then take any program for $\phi$, prepend the interpreter,
and it becomes a $U$-program.  The constant mentioned reflects
the size of such an interpreter.}

Some strings have very short descriptions: a string of a trillion
zeros may be produced by a short program.  Others require descriptions
as long as the strings themselves, for instance a million digit
binary string obtained from a physical random bit generation device.\footnote{
Unless you are rather lucky.}
A recurrent theme in
Kolmogorov complexity is that there are never enough descriptions
to go around so as to give short descriptions to most objects.  
In the case where both the objects and their descriptions are binary
strings, we have the following well-known result that the probability
we can save more than a constant number of bits in compressing 
randomly selected strings is zero.
\begin{fact}[Incompressibility {\cite[\S 2.2]{Li:1997}}]
\label{fact:incompressibility}
Suppose $g : \N \rightarrow \N$ is an integer function with $g(n) > 0$ and
$g \in \omega(1)$,
that is, $\lim_{n \rightarrow \infty} g(n) = \infty$.
Let $x$ be a string chosen \emph{uniformly at random}.
Then almost surely:
\begin{align}
C_\phi(x) \geq \strlen{x} - g\bigl(\strlen{x}\bigr) \label{e:incompressibility}
\end{align}
\end{fact}

\reffact{fact:incompressibility} implies, for example, that one cannot 
devise a coding system that
compresses strings by even $\log \log n$ or $\alpha^{-1}(n,n)$ 
(inverse Ackermann) bits with nonzero probability.
The proof of \reffact{fact:incompressibility} uses counting arguments
only, with no appeal to computability of the description
system.\footnote{
There are $2^{n-g(n)+1}-1$ descriptions of length
at most $n-g(n)$, and $2^{n+1}-1$ strings of length at most $n$.
Therefore the fraction of strings compressible by $g(n)$ bits is at most 
$\frac{2^{n-g(n)+1}-1}{2^{n+1}-1}$,
which behaves in the limit as $2^{-g(n)}$.
If $g \in w(1)$ this value vanishes as $n \rightarrow \infty$,
so $C_\phi(x) \geq \strlen{x} - g(\strlen{x})$ almost surely.
}
Therefore the inequality (\ref{e:incompressibility}) applies to
\emph{any} description system $\phi$, even description systems
that are not computable.  For example, \reffact{fact:incompressibility}
even applies if we permit ourselves to use an infinite, not computably enumerable
library as we described in \refsec{s:platonic}.
However, it does not apply in the case where there is a nonuniform
distribution, as in problem domains where $H < 1$.

In the remainder of this paper we shall assume
compiled programs are incompressible in the sense of
\reffact{fact:incompressibility}.

\begin{prop}
\label{prop:programsincomp}
Compiled C programs on existing major architectures are almost surely
Kolmogorov incompressible.
\end{prop}

Note that ``almost surely \emph{Kolmogorov}
incompressible'' does not imply anything
about the compressibility of \emph{typical} compiled programs for
a problem domain.  Rather, it means that if one
chooses a valid compiled program uniformly at random, with probability 1
it cannot be replaced by a shorter program with the same behaviour.
In subsequent sections we investigate problem domains where
there \emph{is} a nonuniform distribution on programs,
i.e., $\Hdomain < 1$, where the situation is rosier.

We sketch a proof of \refprop{prop:programsincomp}, showing that the
number of distinct behaviours described by compiled programs of $s$ bits
grows as $\sim 2^s$ on current machines, which implies compiled programs are 
almost surely (Kolmogorov) incompressible.
The C
language has the useful ability to incorporate chunks of binary
data in a program.  For example, the binary string $z = 01101001 11011010$
may be encoded by the C declaration
\begin{align*}
\mathsf{unsigned}~\mathsf{char}~\mathsf{z}[2] = \{ \mathsf{0x69}, \mathsf{0xda} \};
\end{align*}
\noindent
Moreover, such arrays are laid out as contiguous binary data in the
compiled program, so that a binary string of length $m$ bytes requires
exactly $m$ bytes in the compiled program.  
We can package such an
array with a short program of constant size that reads the
binary string from memory and outputs it to the console. 
Every binary string of $m$ bytes may be encoded by such a compiled
program of size at most $c + m$ bytes, where $c$ is a constant 
representing the overhead of a read-print loop.  Every such program
yields a unique behaviour, so the number of distinct behaviours of
compiled programs of $s$ bits is $\sim 2^s$.
We can then adapt the argument used to prove
\reffact{fact:incompressibility}, replacing strings by compiled programs,
which shows compiled C programs are almost surely incompressible.


Note that uncompiled programs are \emph{highly} compressible.
For example, C language source code may not contain certain bytes
(e.g., control characters) such as the null character $0x00$.
This means they can be compressed by a factor of 
(at least) $\frac{1}{256} \sim 0.39\%$.
Restricting our attention to \emph{compiled}
programs is crucial.\footnote{An alternative would be to
deal with \emph{indices} of programs in the usual sense
of computability theory, where we equate a program with its
position in some effective enumeration of valid source-language
programs.
However, working with compiled programs has the additional benefit
of brushing aside issues such as identifier lengths in
source code, which tend to be unnecessarily long to
aid readability.
}

%% file: coding.tex
\subsection{Coding of references}

We need some rudimentary accounting of what we gain and lose
by use of the library: we save some 
by using a library component, at the cost of having to refer
to it.  Let us first consider the cost of referring to
components.

We presume that unique identifiers are assigned 
to library components; we call these \emph{codewords}.
Let $c(n)$ be the binary codeword for the $n^{\mathit{th}}$ library
component, and $\strlen{c(n)}$ its length.
Optimal strategies such as Shannon-Fano or Huffman codes assign
shortest codewords to the most frequently
needed components.  Since our library is sorted in order of use
frequency (\refsec{s:ordering}), we may presume that 
$\strlen{c(n)} \leq \strlen{c(n+1)}$, i.e.,
codeword lengths are nondecreasing as we go down the list of components.

In what follows we want to make asymptotic arguments, and
fixing an identifier size (e.g., 64 bits)
would lead to wildly wrong conclusions.\footnote{If we fix memory addresses 
to be representable in 64 bits,
then the time to search an acyclic linked list is $O(1)$ since there are at 
most $2^{64}$ steps the algorithm must go through.
}
Instead we require that the identifier size grows with the number
of components, albeit slowly.  That $\strlen{c(n)} \geq \log_2 n$ 
follows from the pigeonhole principle.
\ifverbose{
\begin{prop}
\label{prop:idlength}
$\strlen{c(n)} \geq \log_2 n$.
\end{prop}
\begin{proof}
Were $\strlen{c(n)} < \log_2 n$, by pigeonhole there would be two components
with the same codeword, contradicting uniqueness.
\qed
\end{proof}
}
Having identifiers of length only $\log_2 n$ leads to difficulties,
because they are not \emph{uniquely decodable}.  That is, if I
am presented with a string of such identifiers I have no way to tell
where one identifier stops and the next starts.  (This does not arise
in current architectures because of fixed word size, but as we said,
care is needed in asymptotic arguments).
A more accurate requirement is the following, which draws on
Kraft's inequality that uniquely decodable codes must satisfy
$\sum_{n=1}^\infty 2^{- \strlen{c(n)}} \leq 1$.

\begin{prop}
\label{prop:logplus}
For identifiers to be uniquely decodable, 
\begin{align*}
\strlen{c(n)} \geq \log^+ n
\end{align*}
where $\log^+ n = \log n + \log \log n + \log \log \log n + \cdots$
and the sum is taken only over the positive real terms.
\ifverbose{\footnote{The notation $\log^\ast x$ is more standard.
However $\log^\ast$ is also widely used
for the \emph{product} of iterated logarithms
$\log n \log \log n \cdots$ that arises in the
context of Zipf's law \cite{Powers:ACL:1998};
$\log^\ast x$ is also commonly used in
algorithm analysis to mean the inverse of a power tower.
Hence we adopt the notation $\log^+ x$ which we hope is
slightly less likely to confuse.
}
}
\end{prop}
We omit the proof; see e.g., \cite[\S 2.2.4]{Rissanen:1989} or
\cite[\S 1.11.2]{Li:1997} (in particular problem 1.11.13).

%% file: newderiv.tex
\subsection{Derivation of reuse rate bound}

\label{s:newderiv}

We now derive an asymptotic upper bound on the rates $\lambda(n)$
at which library components may be reused.  We do this under
the assumption that each time a library component is used in
a program, the same identifier is used to refer to it, i.e.,
there is no \emph{recoding of identifiers}.\footnote{
There are two reasons for this assumption.  (1) On the architectures
from which we collect empirical data, there is no recoding of
identifiers in programs.
(2) The reason one
might want to recode identifiers is to save
space by introducing shorter aliases for components for use
within the program, after the initial reference.
However, this only saves space if a component is more likely to be used
again given it is used once.  While this is intuitively true of
real programs, it is false under a maximum entropy assumption
(\refsec{s:entropy}): in an encoding that maximizes entropy,
the sequence of identifiers in a program behaves statistically as
if independent and identically distributed. 
}
Our argument follows standard lines \cite{Powers:ACL:1998}
but adapted to coding of library references under the model laid 
out in \refsec{s:model}.

\begin{thm}
\label{thm:lambdabound}
Without recoding of identifiers, the asymptotic reuse rates $\lambda(n)$
must satisfy $\lambda(n) \prec (n \log n \log^+ n)^{-1}$.
\end{thm}

\begin{proof}
We count the size of the references to library components within
compressed programs (i.e., those written with use of a library).
Consider programs of length at most $s$.  As $s \rightarrow \infty$, the
expected number of occurrences
of the $n^{\mathit{th}}$ component tends to $\lambda(n) s + o(s)$ under
\refassume{ass:meanlambda}.  Referring to
the $n^{\mathit{th}}$ component requires at least $\log^+ n$ bits (\refprop{prop:logplus}).
We need only
consider components whose identifier length is less than $s$,
since identifiers longer than the program would not fit.
Therefore we consider only up to component number $2^s$ since
$\log^+ 2^s \geq s$.

The expected total size of all the references to components is then at least:
\begin{align*}
\sum_{n=1}^{2^s} \underbrace{\bigl( \lambda(n) s + o(s)\bigr)}_{\text{\# refs}} \underbrace{\log^+ n}_{\text{ref size}}
\end{align*}

The references to components are contained within the program, and
therefore their total size must be less than $s$, the size of the program.
Therefore we have an inequality:\footnote{
\refineqn{e:balancerefs} becomes an equation if we consider programs to
consist solely of a sequence of component references, with
no control flow or other distractions.  This is
possible by building components and programs from combinators,
which can be made self-delimiting \cite[\S 3.2]{Li:1997}.
This provides a theoretically
elegant framework, if not entirely intuitive.
}
\begin{align}
\label{e:balancerefs}
\sum_{n=1}^{2^s} \bigl(\lambda(n) s + o(s)\bigr) \log^+ n &\leq s
\end{align}
Dividing through by $s$ and taking the limit as $s \rightarrow \infty$,
\begin{align}
\lim_{s \rightarrow \infty} \sum_{n=1}^{2^s} \frac{1}{s} \bigl(\lambda(n) s + o(s) \bigr) \log^+ n \leq 1
\end{align}
Since $\lim_{s \rightarrow \infty} \frac{1}{s} o(s) = 0$ by definition,\footnote{Recall that $f \in o(g)$ means $\lim_{x \rightarrow \infty} \frac{f(x)}{g(x)} = 0$.} 
\ifverbose{
\begin{align}
\lim_{s \rightarrow \infty} \sum_{n=1}^{2^s} \lambda(n) \log^+ n \leq 1
\end{align}
or simply
}
\begin{align}
\sum_{n=1}^{\infty} \lambda(n) \log^+ n \leq 1
\label{eqn:eight}
\end{align}
We now consider conditions under which this sum converges.
(\refsec{s:asymptotics} summarizes the asymptotic notations
used here.)
We argue using \refprop{fact:seqbound}, using a diverging
series to bound the terms of \refeqn{eqn:eight}.
The simple argument is to note that the harmonic series diverges,
and therefore the terms of \refeqn{eqn:eight} must grow slower
than this, so $\lambda(n) \log^+ n \prec \frac{1}{n}$, or
$\lambda(n) \prec \frac{1}{n \log^+ n}$.  However,
this bound is quite loose.
\ifverbose{
Tighter bounds follow from the following reasoning.
Suppose $F(x)$ is some rapidly growing invertible function 
\editor{Only need be invertible for $x > 0$}
such as
$e^x$ or $e^{e^x}$.  Then $F^{-1}$ is a slowly growing but unbounded
function, and the series 
\begin{align*}
\sum_{n=0}^\infty \left[ \frac{d}{dx} F^{-1} \right] (n)
\end{align*}
diverges.  (For example, taking $F(x)=e^x$ yields 
$\left[\frac{d}{dx} F^{-1}\right](n) = [\frac{d}{dx} \ln x](n) = \frac{1}{n}$, the harmonic series).
Applying \refprop{fact:seqbound} to \refeqn{eqn:eight}, we have:
\begin{align*}
\lambda(n) \log^+ n \prec \left[ \frac{d}{dx} F^{-1} \right] (n)
\end{align*}
for any rapidly growing invertible function $F$.  Choosing
$F(x) = e^{e^{x}}$ and ignoring constant factors, we obtain:
}
\ifnotverbose{
A more slowly diverging series is $\sum_n \frac{1}{n \log n}$.  Using this,
}
\begin{align*}
\lambda(n) \log^+ n &\prec \frac{1}{n \log n}
\end{align*}
or,
\begin{align}
\boxed{\lambda(n) \prec \frac{1}{n \log n \log^+ n}}
\end{align}
This completes the proof.
\qed
\end{proof}

\ifverbose{
A slightly weaker but simpler bound is 
\begin{align*}
\lambda(n) \prec \frac{1}{n \log^2 n}
\end{align*}
obtained by taking only the first term of $\log^+ n$.
}

The bound of \refthm{thm:lambdabound} is not tight.  No tightest
bound is possible using this line of argument
since there is no slowest diverging sequence with which
to bound a convergent sequence, a classical result due to Niels Abel.\ifverbose{\footnote{
By choosing (say) $F(x)=e^{e^{e^x}}$, we could obtain the bound:
\begin{align*}
\lambda(n) \prec \frac{1}{n \log n \log \log n \log^+ n}
\end{align*}
Even tighter bounds are possible by choosing $F$ to be fast-growing
functions such as the power tower, Ackermann's or the busy beaver function
(using first difference, rather than continuous derivative).
For example, the limiting case of taking $F$ to be iterated exponentials
gives a bound $\lambda(n) \prec (n \log^\ast n \log^+ n)^{-1}$
where $\log^\ast n = \log n \log \log n \log \log \log n \cdots$
and the product is taken only over positive real terms (see, e.g.,
\cite{Powers:ACL:1998}).
}}
However, the bound is tight to within a factor $n^\epsilon$ for any $\epsilon > 0$.
\ifverbose{
That is,
\begin{align*}
\lambda(n) \prec \frac{1}{n^{1+\epsilon} \log n \log^+ n} \preceq \frac{1}{n \log n \log^+ n}
\end{align*}
is a valid bound only if $\epsilon = 0$.
}

\tinysection{Entropy maximization and Zipf's Law.}
\refthm{thm:lambdabound} provides an upper bound on $\lambda(n)$, but it could well be the
case that $\lambda(n) \sim \frac{1}{n^3}$, for example.  Why do the
curves we see in practice (e.g., \reffig{f:rates})
hug the bound of \refthm{thm:lambdabound}?
We believe the answer to why we observe $\lambda(n) \approx \frac{1}{n}$ 
is due to the tendency of libraries to evolve so that 
programmers can write as little code as possible, which in turn
implies evolution toward maximum entropy in compiled code
(\refprin{prin:entropy}).  
The entropy rate of
component references is maximized when $\lambda(n) \approx \frac{1}{n}$
(see, e.g., \cite{Harremoes:E:2001}).

%% file: uniform.tex
\subsection{The uniform case: $\Hdomain=1$}

The uniform case of $\Hdomain=1$, in which every program 
is equally likely to be implemented, reduces the scenario
to classical Kolmogorov complexity with a uniform distribution on
programs.  It has some
surprising properties that suggest $\Hdomain=1$ to be an unlikely scenario
for real problem domains.

Our first result concerns the number of library components we might expect
to use in a program.
Let $N(s)$ be a random variable indicating for a program of uncompressed
size $s$ the number of components whose use reduces program size.
Surprisingly, as program size increases the
expected number of components that reduce program size is bounded
above by a constant.

\begin{thm}
If $\Hdomain=1$ there exists a constant $n_{\mathsf{crit}}$ independent of
program size $s$ such that $N(s) \leq n_{\mathsf{crit}}$ almost surely. 
\end{thm}

\begin{proof}
Suppose each component used saved at least 1 bit.  If
$\lim_{s \rightarrow \infty} E[N(s)]$ were unbounded, use of the
library could compress random programs by an unbounded amount,
contradicting incompressibility (\reffact{fact:incompressibility}). 
\qed
\end{proof}

This has a simple corollary concerning the potential for code reuse.
\begin{cor}
When $\Hdomain=1$ the expected proportion of a program that can be reused
from libraries tends to zero as program size increases.
\end{cor}

\label{s:paradox}
Because of these results, the case $H=1$ is somewhat uninteresting
and does not seem to model real life, where we know libraries are
useful and let us reduce the size of programs.
In the next sections we examine the more interesting case of
$0 < \Hdomain < 1$, where we can compress
programs, even ones that are (Kolmogorov) incompressible,
by use of a library.  

%% file: nonuniform.tex
\subsection{The nonuniform case: $0 < \Hdomain < 1$}

\label{s:nonuniform}
More interesting than the uniform case is the situation when
$0 < \Hdomain < 1$, which implies a nonuniform distribution on
programs.  This models problem domains that have some potential
for code reuse, and libraries are of central importance in
reducing program size.
\ignore{***
In such problem domains libraries let us compress\footnote{In the information theory
sense.} the incompressible.\footnote{In the Kolmogorov complexity sense.}
}
Recall from \refsec{s:diversity} that we can expect to compress programs
in such domains from uncompressed size $s$ to at best $\Hdomain s$ by use of
a library.  A standard result from information theory can be adapted
to show this bound is achievable, at least in a theoretical sense.

\begin{claim}
\label{claim:achieve}
There exists a library with which uncompressed programs of size $s$ can
be compressed to expected size $\sim \Hdomain s$.
\end{claim}

The proof of this is not particularly illustrative
and we banish it to a footnote.\footnote{
\emph{Proof.}
We first describe an encoding that compresses programs to achieve an
expected size $\Hdomain s$, and then explain how to construct the library.
Recall the Shannon-Fano code \cite[\S 1.11]{Li:1997} allows a finite distribution
with entropy $H$ to be coded so that the expected codeword length is
$\leq H+1$.  We adapt this as follows.  For each
$s_0 \in \N$, we produce a Shannon-Fano codebook for all programs of
length $\leq s_0$ achieving average codeword size
$\leq H(p_{s_0}) + 1$ for the distribution $p_{s_0}$ (\refsec{s:diversity}).
By definition $H(p_{s_0}) \leq \Hdomain s$ almost surely, so this achieves
a compression ratio of $\Hdomain$ almost surely for each $s_0$ as $s_0 \rightarrow \infty$.
To combine all the codebooks into one, we preface a compressed program
with an encoding of its uncompressed length, which we use to select the 
appropriate codebook.  This can be done by adding to each codeword $c + 2 \log s$ bits
for some constant $c$, which is negligible with respect to $\Hdomain s$ when
$\Hdomain > 0$.  Therefore this encoding achieves expected program
size $\sim \Hdomain s$.  We use the codebook as the library: each 
component identifier is a Shannon-Fano code, each component is a
program.  Note that the reuse rates vanish for this
construction, i.e., $\lambda(n) \rightarrow 0$ as $s_0 \rightarrow \infty$,
and so the bound of \refthm{thm:lambdabound} is trivially satisfied.
\qed
}
The gist is to place every possible program into the
library as a ``component,'' but ordering them so that the most likely programs for the
problem domain come soonest in the library order and thus are assigned
the shortest codewords.  This is a wildly impractical construction but
demonstrates the claim.  In practice we decompose software into
reusable chunks that we put in libraries; that reusable chunks exist
suggests an ergodic property (see \refsec{s:AEP}).

Unlike the situation of $\Hdomain=1$ where the number of components
useful for a program was at most a constant, when $0 < \Hdomain < 1$ we 
have a much more pleasing situation: the number of useful
components increases steadily as we increase program size.

\subsubsection{The incompleteness of libraries}
\label{s:incompleteness}
Under reasonable assumptions we prove that no finite library
can be complete: there are always more components we can add to
the library that will allow us increase reuse and make programs
shorter.
To make this work we need to settle a subtle
interplay between the Kolmogorov complexity notion of compressibility
(there is a short\-er program doing the same thing)
and the information theoretic notion of compressibility
(low entropy over an \emph{ensemble} of programs).
\ifverbose{
When we defined a problem domain in \refsec{s:distribution},
we did so by stipulating a family of probability distributions
on programs.  From this we derived a bound $\Hdomain$ on the
entropy rate of programs in a problem domain, and said that
we could compress programs to an expected factor of at most $\Hdomain$.
}
Now because we defined probability distributions on programs
(rather than behaviours), we run into the possibility that
the probability distribution might weight heavily
programs that are \emph{Kolmogorov} compressible, i.e.,
the distribution might prefer programs $w$ with $\strlen{w} >> C(w)$.
For example, a problem domain might have programs that are usually
compressible to half their size not
because the probability distribution focuses on a particular
class of problems, but because we artificially defined $p_{s_0}$ to
select only those programs that are twice as large as they might be
(for example, we might pad every likely program with many $\mathsf{nop}$
instructions.)
To avoid this difficulty we require the distributions be \emph{honest}
in the following sense.
\begin{defn}[Honesty]
\label{defn:honesty}
We say the distributions $p_{s_0}$ for a problem domain are
\emph{honest} if the programs are 
Kolmogorov incompressible.  Specifically,
\begin{align}
\label{e:honesty}
E\left[ \frac{C(w)}{\strlen{w}} \right] \rightarrow 1 & ~~~~\text{as } s_0 \rightarrow \infty
\end{align}
where the expectation is taken over the distributions $p_{s_0}$.
This requires that the probability distribution does
not artificially prefer verbose programs with $\strlen{w} >> C(w)$.
\end{defn}

If the distribution for a problem domain is honest and has $\Hdomain < 1$, the 
programs are expected to be \emph{information-theoretically compressible}
by use of a library, but not
\emph{Kolmogorov compressible}.  
In other words, our ability to compress programs
is due to a ``focus'' on a class of problems of interest to the domain,
not just an artificial selection of overly-verbose programs.

Inspired by Euclid's proof that there are infinitely many primes,
with the honesty assumption we can prove there are infinitely many
reusable software components that make programs shorter.

First we need two smaller pieces of the puzzle. 
\ignore{***, which proves a `concentration of
measure' result, that when $\Hdomain > 0$, 
almost all the programs with $\strlen{w} \leq s_0$
are of size nearly $\sim \Hdomain s_0$ or greater.
***}
\begin{lem}
\label{lem:concentration}
If $\Hdomain > 0$ then for any finite $k$, 
$Pr(\strlen{w} \leq k) \rightarrow 0$ as $s_0 \rightarrow \infty$.
\end{lem}
\begin{proof}
We know from definition of $\Hdomain$ that $H(p_{s_0}) = \Hdomain s_0$
infinitely many times as $s_0 \rightarrow \infty$ (\refsec{s:diversity}).  
Consider how probability 
must be distributed among programs of different lengths to account for 
this much entropy.  We try to account for as much
entropy as we can by short programs, setting a uniform distribution
$p(w) = \frac{1}{2^{\Hdomain s_0}}$ on the first $2^{\Hdomain s_0}$ programs---
this is the fewest number of programs that would produce this much entropy.
To programs of length $\leq k$ we can account for
\begin{align*}
\sum_{i=0}^k 2^i \cdot \left( - \frac{1}{2^{\Hdomain s_0}} \log \frac{1}{2^{\Hdomain s_0}} \right)
\sim 2^{k+1-\Hdomain s_0}\Hdomain s_0
\end{align*}
bits of entropy.  But as $s_0 \rightarrow \infty$, 
$2^{k+1-\Hdomain s_0}\Hdomain s_0 \rightarrow 0$ so we can account for
none of the entropy by programs of length $\leq k$.  Therefore
$Pr(\strlen{w} \leq k) \rightarrow 0$ as $s_0 \rightarrow \infty$.
\end{proof}

\begin{lem}
\label{lem:recipzero}
If $\Hdomain > 0$ then $E\left[ \frac{1}{\strlen{w}} \right] \rightarrow 0$ as
$s_0 \rightarrow \infty$.
\end{lem}

\begin{proof}
Suppose $E\left[ \frac{1}{\strlen{w}} \right] = c$ for some $c > 0$.
Then there would be a finite probability that $\strlen{w} \leq c^{-1}$
as $s_0 \rightarrow \infty$, contradicting \reflem{lem:concentration}.
\end{proof}

Now we are ready for the main theorem, which proves no finite library
can be ``complete'' in the sense of achieving a compression ratio of
$\Hdomain$ when $0 < \Hdomain < 1$.

\begin{thm}[Library Incompleteness]
\label{thm:incompleteness}
If a problem domain has $0 < \Hdomain < 1$ and 
honest distributions (\refdefn{defn:honesty}), no
finite library can achieve an asymptotic compression ratio of $\Hdomain$.
\end{thm}

\begin{proof}
Suppose a finite library of components achieves
a compression factor $1 - \epsilon$, with optimal compression when
$1-\epsilon = H$.
Call the programming language $\phi$
and the library $L$.  We can write an interpreter for $\phi$ that
incorporates the library $L$; since the library is finite this is
a finite program.  We call the resulting machine model $\phi+L$.
Consider Kolmogorov complexity for this machine, writing
$C_{\phi +L}(w)$ for the size of the smallest $\phi$-program using $L$ that
has the same behaviour as $w$.  Saying the machine $\phi+L$ achieves the
compression factor $1-\epsilon$ implies
\begin{align}
E\left[ \frac{C_{\phi+L}(w)}{\strlen{w}} \right] = 1 - \epsilon
\end{align}
From the invariance theorem of Kolmogorov complexity (\reffact{fact:invariance})
we have that there exists a constant $c$ such that
\begin{align}
\label{e:zz}
C(w) \leq C_{\phi+L}(w) + c
\end{align}
\noindent
for every program $w$.  Dividing through by $\strlen{w}$ and taking expectation,
\begin{align}
\label{e:zjk}
E\left[ \frac{C(w)}{\strlen{w}} \right] \leq \underbrace{E\left[ \frac{C_{\phi+L}(w)}{\strlen{w}}\right]}_{= 1 - \epsilon} + E\left[ \frac{c}{\strlen{w}} \right] 
\end{align}
From honesty $E\left[ \frac{C(w)}{\strlen{w}} \right] \rightarrow 1$,
and from \reflem{lem:recipzero}
we have
$E\left[\frac{c}{\strlen{w}}\right] \rightarrow 0$.  Therefore (\ref{e:zjk}) is, in the limit as $s_0 \rightarrow \infty$:
\begin{align*}
1 \leq (1 - \epsilon) + 0
\end{align*}
For this inequality to hold,
$\epsilon \rightarrow 0$ for any finite library.  Therefore
no finite library can achieve an asymptotic compression ratio $< 1$
when the distributions are honest.
\qed
\end{proof}



\refclaim{claim:achieve} showed that an infinite library can achieve
expected size $\sim \Hdomain s$; \refthm{thm:incompleteness} shows
that no finite library can.  Therefore only infinite libraries can
compress programs of size $s$ to expected size $\Hdomain s$.  
However, this is an asymptotic argument; if we restrict ourselves
to programs of size $\leq s_0$ for some fixed $s_0$, finite
libraries can approach a compression ratio of $\Hdomain s$ by including
more and more components.
\ignore{***\footnote{
This can be modelled by considering truncated versions of the entropy
series
\begin{align}
\label{s:entropyseries}
\sum_{n=1}^\infty H(\lambda)
\end{align}
\noindent
where $H(\lambda) = - \lambda \log \lambda$ is the usual definition of
entropy.  The convergence of the sum \refeqn{s:entropyseries} 
is very slow, suggesting there are always more components to
add to the library that add practical savings in expected codeword size.
\editor{It would be nice to show a plot!  The ``very slow'' part
comes because we choose the series to be just on the cusp of a
divergent series?}
}
***}
Doug Gregor suggested calling
\refthm{thm:incompleteness}
the \emph{Full Employment Theorem for Library Writers}, after 
Andrew Appel's boon to compiler writers.
\refthm{thm:incompleteness} has a straightforward implication:
no finite library can be complete; there are always more useful
components to add.  In practice we have a tradeoff between the
utility of larger libraries and the economic cost of
producing them; this suggests the importance of designing libraries
for extensibility. 


A minor change to the above proof yields a similar but slightly stronger
result.
\begin{cor}
\label{cor:noce}
If a problem domain has $0 < \Hdomain < 1$ and honest distributions, no
\emph{computably enumerable} library can achieve a compression ratio of
$\Hdomain$.
\end{cor}
\begin{proof}
Repeat the proof of \refthm{thm:incompleteness}, replacing ``finite library''
with ``c.e. library.''  In particular the choice of a c.e. library guarantees
that the interpreter for $\phi+L$ is a finite program: whenever a library
subroutine is required, it can be generated from the program enumerating the
library.
\qed
\end{proof}

We may casually equate ``not computably enumerable'' with ``requires
human creativity.''  \refcor{cor:noce} indicates that the process of
discovering new and useful library components is not a process that can
can be fully automated.


%% file: size.tex
\subsubsection{Size of library components.}
\label{s:size}
We now consider how big library components might be.
If we want to achieve the strong reuse vision of ``programming by wiring
together large  components,'' this suggests that components ought to be
quite large compared to the wiring.  The following theorem sheds
light on the conditions when this is possible.

Let $S(n)$ denote the expected amount of code (in bits) saved
per use of the $n^\mathit{th}$ component.  We consider
the case when $\lambda(n) \sim \frac{1}{n \strlen{c(n)} f(n)}$,
where $\strlen{c(n)}$ is the codeword (identifier) length,
and $f(n)$ is a function $f \in o(n^\epsilon)$ for $\epsilon > 0$
that ensures convergence (cf. \refsec{s:newderiv}).
This coincides with a Zipf-style law as observed in practice
(\reffig{f:rates}).

\begin{thm}
\label{thm:size}
If a library achieves a compression factor of $\Hdomain > 0$ in an
honest problem domain, then 
$S(n) \sim \frac{1-\Hdomain}{\Hdomain} \cdot o(n^\epsilon)$
for any $\epsilon > 0$.
\end{thm}
\begin{proof}
Summing over all components, the total code saved is:
\begin{align}
\sum_{n=1}^\infty \underbrace{\bigl(\lambda(n) \Hdomain s + o(\Hdomain s)\bigr)}_{\text{expected \# uses}} ~~\cdot \underbrace{S(n)}_{\text{savings per use}} = \underbrace{(1-\Hdomain) s}_{\text{total savings}}
\end{align}
\noindent
Dividing through by $\Hdomain s$ and taking the limit as $s \rightarrow \infty$,
and substituting $\lambda(n) \sim \frac{1}{n \strlen{c(n)} f(n)}$,
\begin{align*}
\sum_{n=1}^\infty \frac{1}{n \strlen{c(n)} f(n)} S(n) = \frac{1-\Hdomain}{\Hdomain}
\end{align*}
Now if $S(n) \sim n^a$ for some constant $a > 0$ then the sum would diverge.  Therefore
$S(n)$ is not polynomial in $n$; in fact for the sum to converge we must
have $S(n) \prec f(n)$ which means $S(n)$ behaves asymptotically as
\begin{align*}
S(n) \sim \frac{1-\Hdomain}{\Hdomain} o(n^\epsilon)
\end{align*}
where $o(n^\epsilon)$ denotes some subpolynomial function.
\qed
\end{proof}

Note that if the components in the library are unique, then
$S(n) \geq \log n$ by pigeonhole.

\begin{figure*}
\begin{centering}
\includegraphics[scale=0.7]{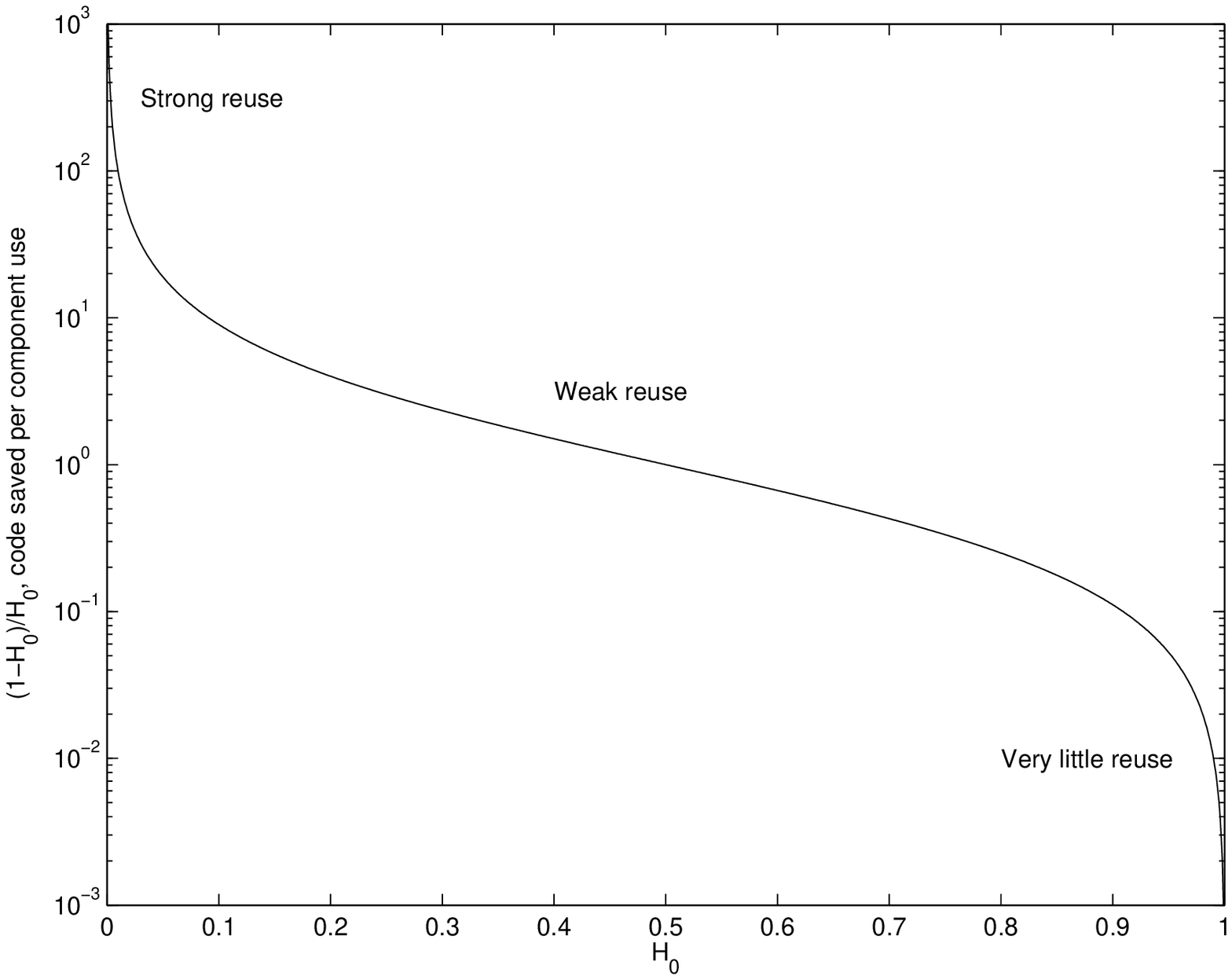}

\end{centering}

\caption{Plot of $\frac{1-\Hdomain}{\Hdomain}$ versus $\Hdomain$, indicating
how much code is saved, proportionately, per component use.  When
$\Hdomain \rightarrow 1$ there is almost no reuse; $\Hdomain \rightarrow 0$
coincides with the ``strong reuse'' ideal of wiring together large
components.  In between is weak reuse, with moderate amounts of code
drawn from libraries.
}
\end{figure*}

\tinysection{Strong reuse?}
The interpretation of \refthm{thm:size} is fairly intuitive.  Roughly
it says the savings we can expect per component are linear
in the size of the component identifier.  Which is to say, we
should expect savings for the $n^\mathit{th}$ component to grow roughly 
as $\log^{+} n$.
This is consistent with findings in the
reuse literature that \emph{small} components are much more likely
to be reused.  
The important factor here is the multiplier $\frac{1-\Hdomain}{\Hdomain}$.
As $\Hdomain \rightarrow 0$, this multiplier becomes arbitrarily large.
This suggests that ``strong reuse'' (\refsec{s:intro}) corresponds
to the region $\Hdomain \rightarrow 0$.  
For example, if programs in a problem domain are thought to be solvable 
by wiring together components that are (say) $1000$ times bigger than the 
wiring itself, this suggests $\frac{1-\Hdomain}{\Hdomain} \approx 10^5$ or
$\Hdomain \approx 0.001$.  The key result is that whether one
is able to achieve strong reuse depends critically on the parameter 
$\Hdomain$ --- which measures how much diversity there is in the
problem domain.

%% file: data.tex
\section{Experimental data collection}

\label{s:data}

Preliminary empirical data was collected from three large Unix
installations.
The problem domain is not particularly well-defined,
but is rather ``the mishmash of things one wants to do on a
typical research Unix machine.''  On the SunOS and Mac OS X machines
we located every
shared object and used the unix commands $\mathsf{nm}$ or
$\mathsf{objdump}$ to obtain a listing of the relocatable symbols
(i.e., references to subroutines in shared libraries).  
For the Linux machine, a more sophisticated approach was used that
involved disassembling every executable object and decoding the
PLT and GOT tables for shared library calls.  For this reason
the Linux data is much more fine-grained and reliable; for example,
our data set for Linux includes the frequencies of all the x86
machine instructions, in addition to almost a half-million subroutines.

\begin{tabular}{lrr}
Operating System & \# Objects & \# Components \\ \hline
Linux (SuSE) & 12136 & 455716 \\
SunOS & 23774 & 110306 \\
Mac OS X & 2334 & 37677
\end{tabular}

We counted the number of references to each component, sorted these
by frequency, and this data is plotted in \reffig{f:rates}.  
The observed counts match nicely the asymptotic prediction made in
\refsec{s:newderiv} (the family of curves $c n^{-1}$ is shown
as dotted lines).
To account for machine instructions, which are not included in the tally
for the Mac OS X and SunOS machines
but constitute by far the most frequently occurring software components,
we started numbering the components for these machines at $n=50$.  Without this
adjustment the rates have a characteristic ``flat top'' and then
rapidly converge to $n^{-1}$ lines; this is an artifact of the
log-log scale.  

The pronounced ``steps'' in the data for large $n$ occur because
there are many rarely-used subroutines with only a few references;
this is typical of such plots (see, e.g., \cite{Powers:ACL:1998}).

\begin{figure*}
\subfigure[Scatter-plot of the number of subroutine references]{
\includegraphics[scale=0.6]{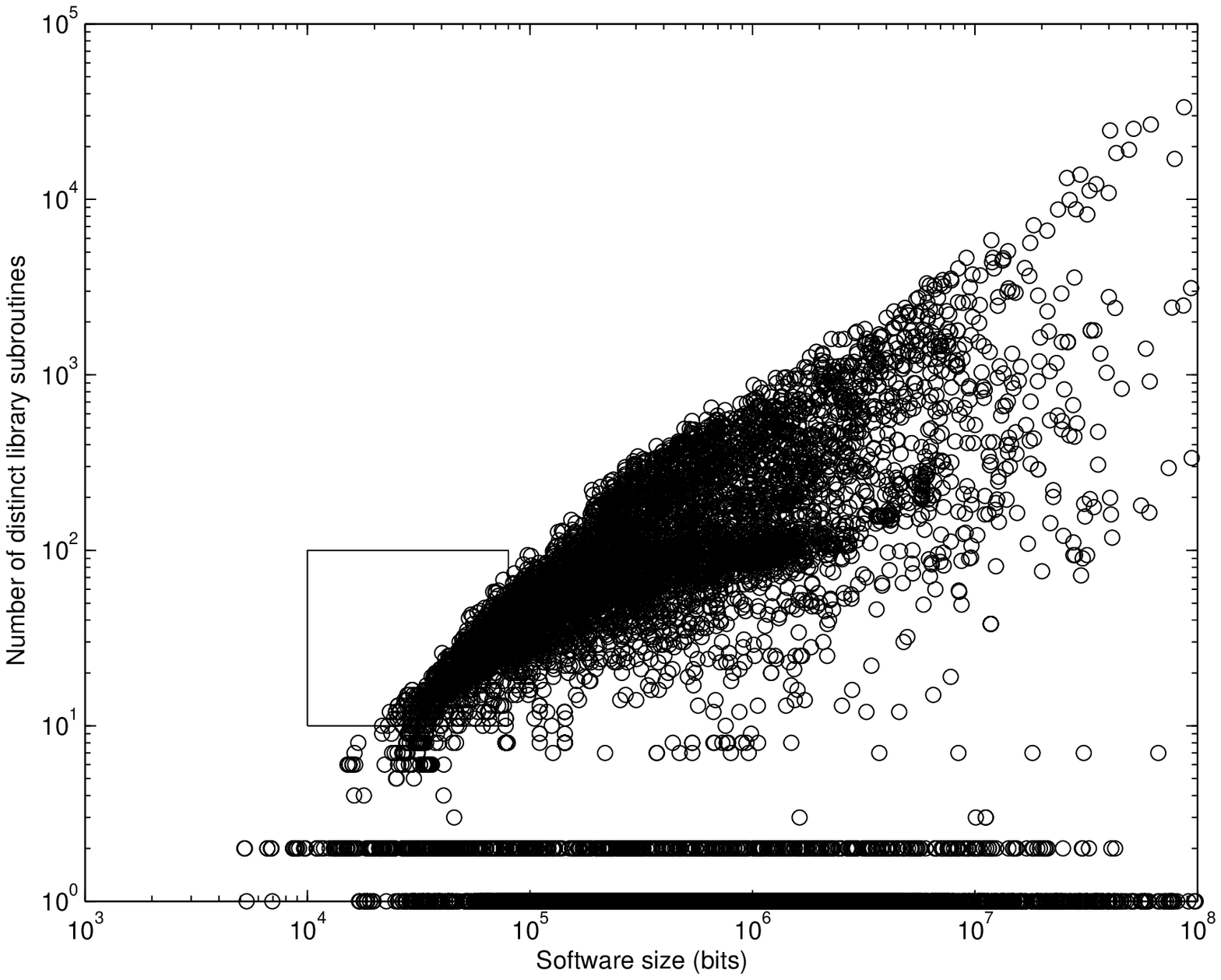}
}
\begin{minipage}[b]{2in}
\subfigure[Distribution histogram]{
\includegraphics[scale=0.4]{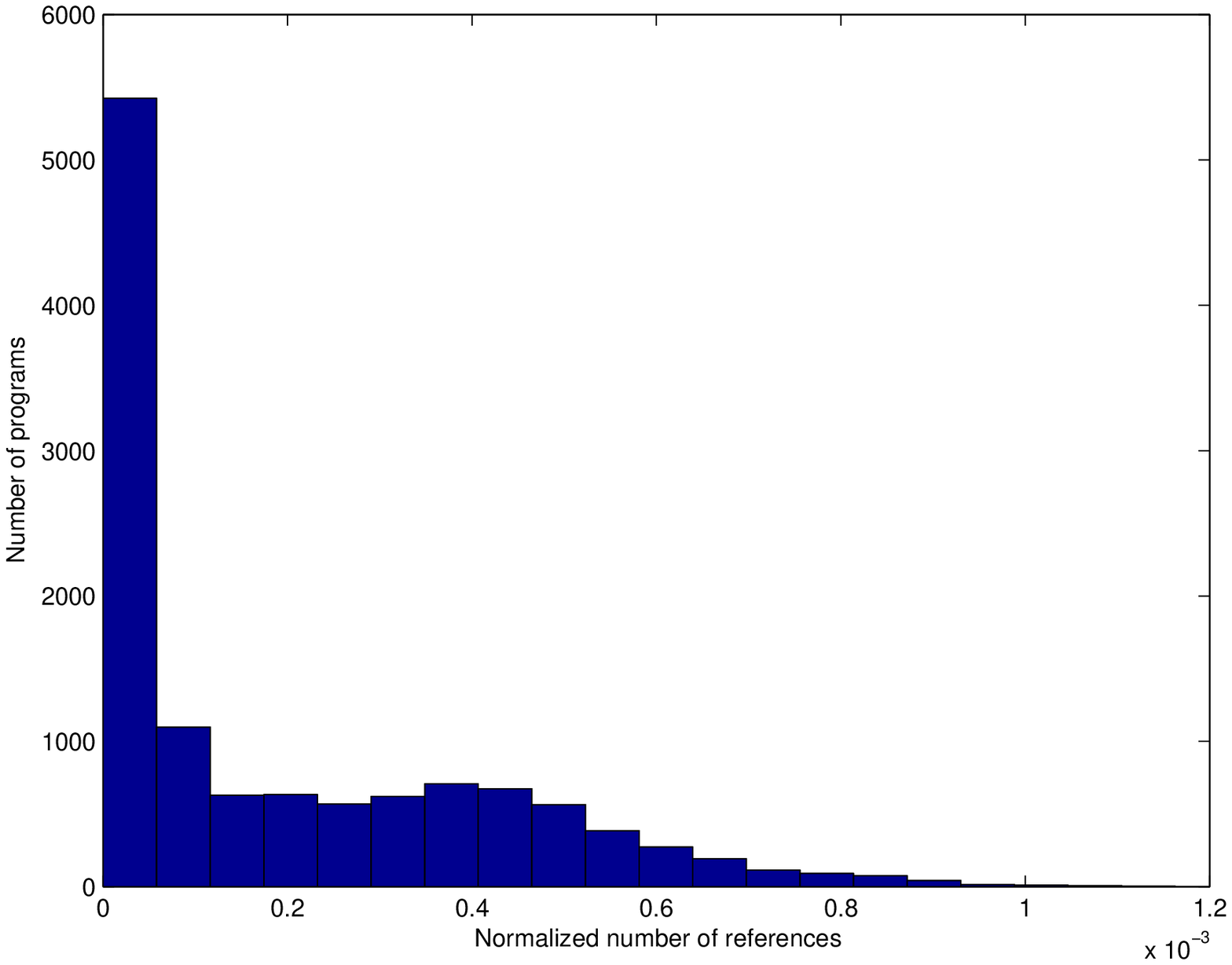}
} 
\subfigure[Distribution for inset box]{
\includegraphics[scale=0.4]{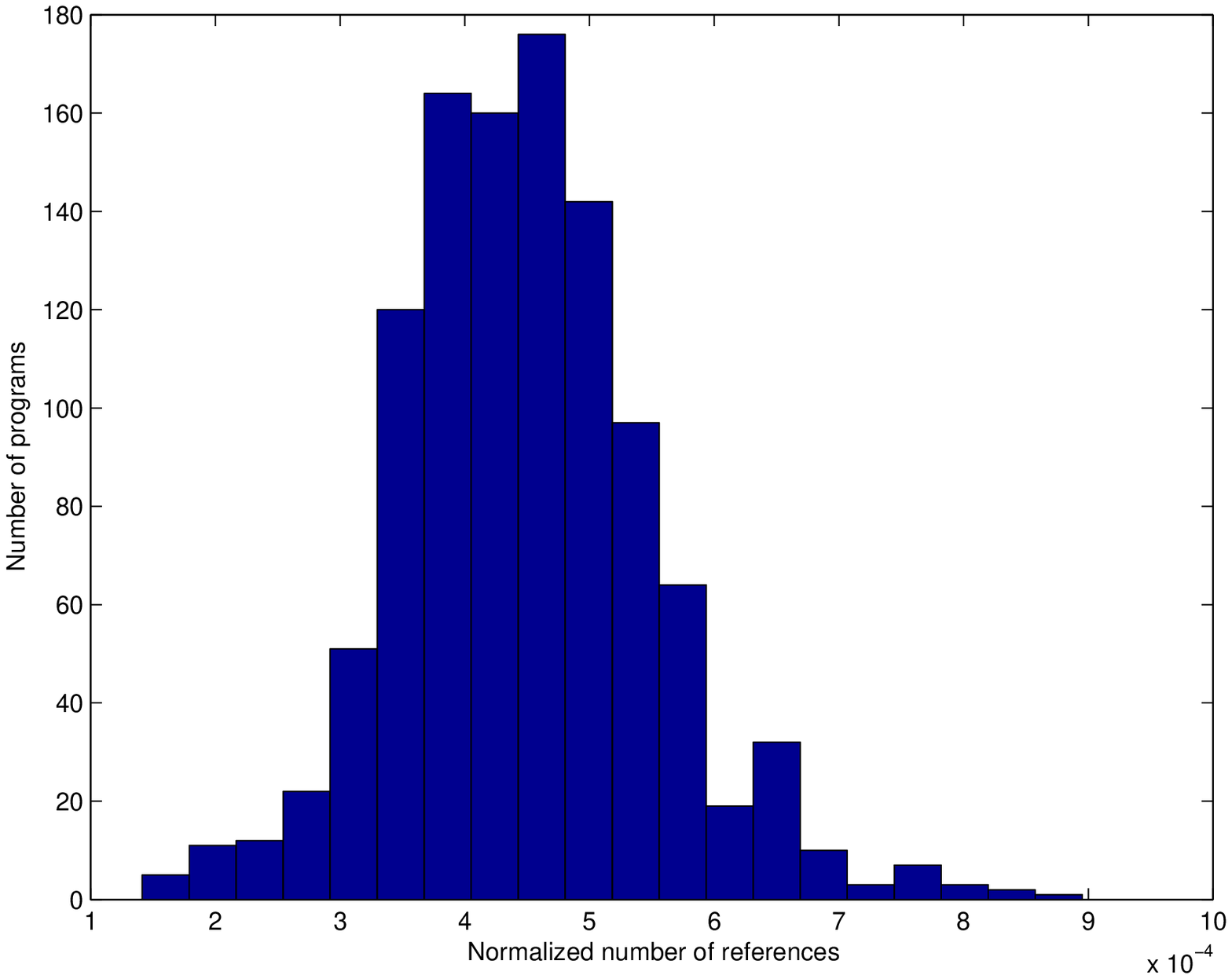}
}
\end{minipage}
\caption{\label{f:erdos}Data suggesting a library analogue of the
Erd{\"o}s-Kac theorem.  (a) A scatter-plot showing the number of
distinct library subroutines used vs. software size for the Linux RPM
data.  (b) Histogram for the number of references, normalized (see text).
(c) Histogram only for the inset box of (a), illustrating an
Erd{\"o}s-Kac-style normal distribution for the number of components used
in software.  Such plots might provide a useful tool for assessing
the extent of reuse vs. ideal predictions from a model.}
\end{figure*}

Another prediction that may follow from our model is that the number 
of distinct components used in a program should approach a normal distribution:
under maximum entropy conditions the use of components is statistically
independent, and so the central limit theorem applies.
This is reminiscent of the
Erd{\"o}s-Kac theorem \cite{Erdos:AJM:1940} that the number of prime factors 
of integers converges to a normal distribution.  \reffig{f:erdos} shows some
preliminary results that support this result, drawn from the
SuSE Linux data.  The number of component references have been
normalized by an estimated variance of $\sigma^2 = c s^2$ where
$s$ is the program size.  Subfigure (c) shows a suggestively
shaped distribution for the inset box of (a), a region where there
is good ``saturation'' of the problem domain with programs.

Our preliminary data demonstrates a Heaps' style law for
vocabulary growth \cite[\S 7.5]{Heaps:1978}: the number of unique 
components encountered
in examining the first $s$ bytes of the corpus grows roughly
as a power law $s^\alpha$ with $\alpha \approx 0.8$.  We have
not found a satisfactory theoretical explanation.




%% file: conclusion.tex
\section{Conclusion}

\label{s:conclusion}
We have developed a theoretical model of reuse libraries that provides
good agreement, we feel, with our intuitions, the literature, and the
preliminary experimental data we have collected on reuse on Unix
machines.  Much of what we have done has served to emphasize the
importance of this one quantity, $\Hdomain$, the entropy rate we
associate with a problem domain.  It determines if we can have
strong reuse ($\Hdomain \rightarrow 0$), or whether we can have
weak reuse $(0 < \Hdomain < 1)$, and how much code we might be able
to reuse from libraries: at most $1-\Hdomain$. 

We have shown that libraries allow us to ``compress the incompressible,''
reducing the size of programs that are Kolmo\-gorov-incompressible
by taking advantage of the commonality exhibited by programs within
a problem domain.  We have also shown that libraries are essentially
incomplete, and there will always be room for more useful components
in any problem domain.  

\ignore{***
Interestingly, the invariance theorem of
Kolmogorov complexity (\refthm{fact:invariance}) implies that
domain-specific \emph{languages}, if assumed to have a fixed and finite 
definition, cannot achieve any asymptotic reduction in program size
for a problem domain.
However, domain-specific \emph{libraries}, if considered ever-growing 
approximations to a ``Platonic library,'' can get arbitrarily close 
to the optimal amount of compression allowed by the parameter $\Hdomain$.
***}

The arguments made here are quite general and might adapt easily
to other description systems, for example, the reuse of
abstractions, lemmas and theorems in mathematical proofs.

%% file: biblio.tex
\bibliography{KEYWORDS,abint,activelib,alg,algebra,alias,arrays,asymptotics,autocomplexity,automata,checking,coalgebra,cocv,coding,combinatorics,compilers,complexity,components,computability,cpp,dsl,extensible,fixpoint,generics,graph,hardware,hashing,hci,historyofcomputing,incremental,java,kolmogorov,lang,lattice,libraries,logic,matrixmult,meta,nonstandard,numanalysis,numtheory,oon,order,ordinals,parallel,parsing,partial,persistent,philosophy,prob,processes,realtime,recurrence,representations,research,reuse,rewrite,safety,security,selection,semantics,settheory,slicing,software,solvers,staging,stats,steiner,stores,subrecursive,tcs,theories,tveldhui,types,verify,writing,zipf}

%% file: asymptotics.tex
\subsection{Asymptotics}

\label{s:asymptotics}
Here we recall briefly some facts and notations concerning 
asymptotic behaviour of functions and series.
For a more detailed exposition we suggest \cite{Graham:1994}.

\tinysection{Asymptotic notations.}
For positive functions $f(n)$ and $g(n)$, we make use of these
notations for asymptotic behaviour:
\begin{align}
f(n) \sim g(n) ~~ & \Longleftrightarrow ~~ \lim_{n \rightarrow \infty} \frac{f(n
)}{g(n)}=1 \\
f(n) \prec g(n) ~~ & \Longleftrightarrow ~~ \lim_{n \rightarrow \infty} \frac{f(n)}{g(n)}=0 \\
f(n) \preceq g(n) ~~ & \Longleftrightarrow ~~ \exists c \in \R ~.~ \lim_{n \rightarrow \infty} \frac{f(n)}{g(n)}=c 
\end{align}
The ``big-O'' style of
notation $f \in o(g)$ is equivalent to $f(n) \prec g(n)$.
When we write $h(n) \sim g(n) + o(n^2)$ we mean
there exists some function $f \in o(n^2)$ such that
$h(n) \sim g(n) + f(n)$.

\tinysection{Series and their convergence.}
A series $\sum_{i=1}^\infty a_i$ is convergent when
$\lim_{N \rightarrow \infty} \sum_{i=1}^N a_i$ exists
in the standard reals; otherwise it is divergent.
The Harmonic series $h_n = \frac{1}{n}$ is divergent,
since $\sum_{i=0}^\infty h_i = 1 + \frac{1}{2} + \frac{1}{3} + \cdots$
fails to converge.

We shall make use of the following key fact for bounding convergent
sequences.
\begin{fact}
\label{fact:seqbound}
Let $a_n,b_n$ be positive sequences.
If $\sum_{n=1}^\infty a_n$ converges and $\sum_{n=1}^\infty b_n$
diverges, then $a_n \prec b_n$.
\end{fact}
\refprop{fact:seqbound} is useful to establish a bound on the asymptotic
growth of a sequence: for example, if $\sum_{n=1}^\infty a_n$ must converge, 
then $a_n \prec \frac{1}{n}$ since the harmonic series diverges.

%% file: libcomp.bbl
\begin{thebibliography}{10}

\bibitem{Ash:1967}
Robert Ash.
\newblock {\em Information Theory}.
\newblock John Wiley \& Sons, New York, 3 edition, 1967.

\bibitem{Balaguer:SEP:2004}
Mark Balaguer.
\newblock Platonism in {Metaphysics}.
\newblock In Edward~N. Zalta, editor, {\em The Stanford Encyclopedia of
  Philosophy}. Summer 2004.

\bibitem{Basili:CACM:1996}
Victor~R. Basili, Lionel~C. Briand, and Walcélio~L. Melo.
\newblock How reuse influences productivity in object-oriented systems.
\newblock {\em Commun. ACM}, 39(10):104--116, 1996.

\bibitem{Ben-David:JCSS:1992}
S.~Ben-David, B.~Chor, O.~Goldreich, and M.~Luby.
\newblock On the theory of average case complexity.
\newblock {\em Journal of Computer and System Sciences}, 44(2):193--219, April
  1992.

\bibitem{Berry:SIGPLAN:1983}
Daniel~M. Berry.
\newblock A new methodology for generating test cases for a programming
  language compiler.
\newblock {\em SIGPLAN Not.}, 18(2):46--56, 1983.

\bibitem{Compton:NATO:1988}
Kevin~J. Compton.
\newblock 0--1 laws in logic and combinatorics.
\newblock In I.~Rival, editor, {\em Proceedings NATO Advanced Study Institute
  on Algorithms and Order}, pages 353--383, Dordrecht, 1988. Reidel.

\bibitem{Cover:1991}
Thomas~M. Cover and Joy~A. Thomas.
\newblock {\em Elements of information theory}.
\newblock Wiley series in telecommunications. John Wiley \& Sons, 1991.

\bibitem{Erdos:AJM:1940}
P.~Erd{\"o}s and M.~Kac.
\newblock The {G}aussian law of errors in the theory of additive number
  theoretic functions.
\newblock {\em Amer. J. Math.}, 62:738--742, 1940.

\bibitem{Frakes:SE:1996}
William~B. Frakes and Christopher~J. Fox.
\newblock Quality improvement using a software reuse failure modes model.
\newblock {\em IEEE Transactions on Software Engineering}, 22(4):274--279,
  April 1996.

\bibitem{Graham:1994}
Ronald~L. Graham, Donald~E. Knuth, and Oren Patashnik.
\newblock {\em Concrete Mathematics: {A} Foundation for Computer Science}.
\newblock Ad{\-d}i{\-s}on-Wes{\-l}ey, Reading, MA, USA, second edition, 1994.

\bibitem{Green:HCI:1989}
T.~R.~G. Green.
\newblock Cognitive dimensions of notations.
\newblock In {\em Proceedings of the HCI'89 Conference on People and Computers
  V}, Cognitive Ergonomics, pages 443--460, 1989.

\bibitem{Griss:IBMSJ:1993}
M.~L. Griss.
\newblock Software reuse: From library to factory.
\newblock {\em IBM Systems Journal}, 32(4):548--566, 1993.

\bibitem{Harremoes:E:2001}
P.~Harrem{\"o}es and F.~Tops\o e.
\newblock Maximum entropy fundamentals.
\newblock {\em Entropy}, 3(3):191--226, 2001.

\bibitem{Heaps:1978}
H.~S. Heaps.
\newblock {\em Information retrieval: theoretical and computational aspects}.
\newblock Academic Press, New York, NY, 1978.

\bibitem{Krueger:CSUR:1992}
Charles~W. Krueger.
\newblock Software reuse.
\newblock {\em ACM Comput. Surv.}, 24(2):131--183, 1992.

\bibitem{Kuck:1978}
D.~Kuck.
\newblock {\em The Structure of Computers and Computations, {Volume 1}}.
\newblock John Wiley and Sons, New York, NY, 1978.

\bibitem{Laemmel:TR:1977}
A.~Laemmel and M.~Shooman.
\newblock Statistical (natural) language theory and computer program
  complexity.
\newblock Technical Report POLY/EE/E0-76-020, Dept of Electrical Engineering
  and Electrophysics, Polytechnic Institute of New York, Brooklyn, August 15
  1977.

\bibitem{Latendresse:IVME:2003}
Mario Latendresse and Marc Feeley.
\newblock Generation of fast interpreters for {Huffman} compressed bytecode.
\newblock In {\em IVME '03: Proceedings of the 2003 workshop on Interpreters,
  virtual machines and emulators}, pages 32--40, New York, NY, USA, 2003. ACM
  Press.

\bibitem{Li:1997}
M.~Li and P.~Vitányi.
\newblock {\em An introduction to Kolmogorov complexity and its applications}.
\newblock Springer-Verlag, New York, 2nd edition, 1997.

\bibitem{Li:HTCS:1990}
M.~Li and P.~M.~B. Vitányi.
\newblock {K}olmogorov complexity and its applications.
\newblock In Jan van Leeuwen, editor, {\em Handbook of Theoretical Computer
  Science}, volume A: Algorithms and Complexity. Elsevier, New York, NY, USA,
  1990.

\bibitem{Li:WWW:2005}
Wentian Li.
\newblock Bibliography on {Zipf's Law}, 2005.
\newblock http://www.nslij-genetics.org/wli/zipf/index.html.

\bibitem{Mohagheghi:ICSE:2004}
Parastoo Mohagheghi, Reidar Conradi, Ole~M. Killi, and Henrik Schwarz.
\newblock An empirical study of software reuse vs. defect-density and
  stability.
\newblock In {\em ICSE '04: Proceedings of the 26th International Conference on
  Software Engineering}, pages 282--292, Washington, DC, USA, 2004. IEEE
  Computer Society.

\bibitem{Poulin:1997}
Jeffrey~S. Poulin.
\newblock {\em Measuring Software Reuse: Princples, Practices, and Economic
  Models}.
\newblock Addison-Wesley, 1997.

\bibitem{Powers:ACL:1998}
David M.~W. Powers.
\newblock Applications and explanations of {Z}ipf's law.
\newblock In Jill Burstein and Claudia Leacock, editors, {\em Proceedings of
  the Joint Conference on New Methods in Language Processing and Computational
  Language Learning}, pages 151--160. Association for Computational
  Linguistics, Somerset, New Jersey, 1998.

\bibitem{Rissanen:1989}
Jorma Rissanen.
\newblock {\em Stochastic Complexity in Statistical Inquiry}, volume~15 of {\em
  Series in Computer Science}.
\newblock World Scientific, 1989.

\bibitem{Ross:1996}
Sheldon~M. Ross.
\newblock {\em Stochastic Processes}.
\newblock John Wiley and Sons; New York, NY, 2nd edition, 1996.

\bibitem{Wortman:1973}
David~Barkley Wortman.
\newblock {\em A study of language directed computer design}.
\newblock PhD thesis, Stanford University, 1973.

\end{thebibliography}
